\documentclass[aps,twocolumn,preprintnumbers,nofootinbib,superscriptaddress]{revtex4-1}
\usepackage{times}
\usepackage{eurosym}
\usepackage{amsmath}
\usepackage{graphicx}
\usepackage{amsfonts}
\usepackage{array}
\usepackage{amsthm}
\usepackage{bm}
\usepackage{supertabular}
\usepackage{subfig}
\usepackage[breaklinks]{hyperref}
\usepackage{graphicx}
\usepackage{caption}
\usepackage[font=small,labelfont=bf,justification=raggedright]{caption}
\usepackage{color}
\usepackage{booktabs}
\usepackage{multirow}
\newcommand{\be}{\begin{equation}}
	\newcommand{\ee}{\end{equation}}
\newcommand{\ba}{\begin{eqnarray}}
	\newcommand{\ea}{\end{eqnarray}}
\newcommand{\bal}{\begin{align}}
	\newcommand{\eal}{\end{align}}

\newcommand{\bw}{\begin{widetext}}
	\newcommand{\ew}{\end{widetext}}

\usepackage[utf8]{inputenc}
\usepackage{amssymb}
\usepackage{amsmath}
\usepackage{amsfonts}
\usepackage{graphicx, float}
\usepackage{graphicx, epsfig}
\usepackage{color}
\usepackage{enumerate}
\newcommand{\beq}{\begin{equation}}
	\newcommand{\eeq}{\end{equation}}
\newcommand{\bea}{\begin{eqnarray}}
	\newcommand{\eea}{\end{eqnarray}}

\begin{document}
	
	\title{Constraining wormhole geometries using the orbit of S2 star and the Event Horizon Telescope}
	
	\author{Kimet Jusufi}
	\email{kimet.jusufi@unite.edu.mk}
	\affiliation{Physics Department, State University of Tetovo, Ilinden Street nn, 1200, Tetovo, North Macedonia}
	\author{Saurabh Kumar}
	\email{sbhkmr1999@gmail.com}
	\affiliation{Department of Physics, Dyal Singh College, University of Delhi, 110003, India}
	\author{Mustapha Azreg-A\"{\i}nou}
	\email{azreg@baskent.edu.tr}
	\affiliation{Engineering Faculty, Ba\c{s}kent University, Ba\u{g}l{\i}ca Campus, 06790-Ankara, Turkey}
	\author{Mubasher Jamil}
	\email{mjamil@sns.nust.edu.pk}
	\affiliation{Institute for Theoretical Physics and Cosmology, Zhejiang University of Technology, Hangzhou 310023 China}
	\affiliation{School of Natural Sciences, National University of Sciences and Technology, Islamabad, 44000, Pakistan}
	
	\author{Qiang Wu}
	\email{wuq@zjut.edu.cn}
	\affiliation{Institute for Theoretical Physics and Cosmology, Zhejiang University of Technology, Hangzhou, 310023 China}
	
	\author{Cosimo Bambi}
	\email{bambi@fudan.edu.cn}
	\affiliation{Center for Field Theory and Particle Physics and Department of Physics, Fudan University, 200438 Shanghai, China}

	\begin{abstract}
		In this paper we study the possibility of having a wormhole (WH) as a candidate for the Sgr A$^\star$ central object and test this idea by constraining their geometry using the motion of S2 star and the reconstructed shadow images. In particular, we consider three WH models, including WHs in Einstein theory, brane-world gravity, and Einstein-Dirac-Maxwell theory. To this end, we have constrained  the WH throat using the motion of S2 star and shown that the flare out condition is satisfied. We also consider the accretion of infalling gas model and study the accretion rate and the intensity of the electromagnetic radiation as well as the shadow images.
	\end{abstract}
	
	\maketitle

	\section{Introduction}
	
	WHs are tube-like structures which may connect, as shortcuts, two or more spatially or temporally separated regions in the spacetime. Geometrically, they are non-singular and traversable  structures which admit a throat. The elusive WH geometry is supported by the Einstein’s theory of general relativity (GR) being one of the exact solutions of Einstein field equations (see \cite{Morris:1988cz,Bambi:2021qfo} for historical reference and reviews). Classically, the existence and stability of WH geometry requires the presence of negative energy density and the violation of the weak energy condition, however other approaches such as  quantum gravity and quantum field theory in curved spacetime minimize such violations near the WH’s throat. In classical GR, type I WHs derived in~\cite{Azreg:2015} are the solutions to the field equations that violate the least the local energy conditions. 
	Besides, WHs have been realized in extended or beyond GR theories as well including Einstein-Gauss-Bonnet \cite{Mehdizadeh:2015jra}, Brans-Dicke \cite{Tretyakova:2015vaa}, scalar-tensor \cite{Bahamonde:2016jqq}, $f(R)$ \cite{Bahamonde:2016ixz}, Braneworld \cite{Tomikawa:2014wxa}, 4D Gauss-Bonnet gravity \cite{Jusufi:2020yus} and in Einstein-Maxwell-Dirac theory \cite{Blazquez-Salcedo:2020czn}. 
	
	Currently, the WH paradigm is one of the leading candidates for the tests of strong gravity and high energy astrophysical phenomenon in the observable universe. For instance, different forms of WHs have been tested to reproduce the result of the recent detection of shadows of an object at the center of galaxy M87~\cite{Azreg:2015, W2}, numerous microquasars with quasiperiodic oscillations \cite{Deligianni:2021ecz} and the gravitational lensing events by compact gravitational sources \cite{Asada:2017vxl}, since observations do not strictly support interpretations via black hole (BH) models solely~\cite{Azreg:2015}.  Consequently, numerous astrophysical tests of traversable WHs have been carried out for distinguishing WHs from BHs to further test the limits of GR and beyond-GR theories. 
	
	In the present context, it would be central to understand how does the dynamics of a star and light around a WH differ from that of a BH. It has been noted in \cite{Kitamura:2012zy} that WHs can cause demagnification of images of background sources unlike BHs, which can constrain exotic spacetime geometries. For a BH, light moves around in a null circular geodesic constituting a photon sphere outside the event horizon, analogously a photon sphere is formed outside the WH’s throat. In order to detect the shadows of any astrophysical WHs by an observer, an optically thin accretion disk of gas surrounding the WH is necessary \cite{Karimov:2020fuj}. However, due to weak gravitational field of WHs compared to BHs, the size of the shadow's boundary is expected to be smaller. Since every WH has a different shadow boundary, the detection of the shadow of M87 galaxy’s central object has already ruled out certain models of WHs~\cite{Bambi:2021qfo}. From a theoretical point of view, the investigation of the shadow of the Sgr A$^\star$ has also ruled out certain types of WHs~\cite{Azreg:2015}. Further, the S-cluster of stars orbits the Sgr A$^\star$ in which one of the stars known as the S2 star experiences an acceleration close to $1.5 \text{ m/s}^2$ which can be tested using both BH and WH backgrounds.
	
	Another improved test of astrophysical detection of WHs is the X-ray analysis of the radiation emitted from the inner regions of the accretion disks surrounding compact objects \cite{Tripathi:2019trz}, where the disk needs to be geometrical thin and optically thick, within the standard framework of Novikov-Thorne model. The reflection spectrum of a BH is reported to be markedly different from a WH provided the accreting gas transfers from one mouth of the WH to the next one. This process produce novel signatures in the emission and reflection spectra. Moreover, after the detection of gravitational wave events due to merger of binary BHs, a new window to gravitational wave astronomy has just opened \cite{Abbott:2016blz}. By looking in the details of the merger and the quasinormal modes detected for each gravitational wave event, it would be possible to distinguish if the merger culminates in a BH or a WH \cite{Konoplya:2018ala}. In the similar vein, the scattering of gravitational waves by WHs and other compact objects might create a stochastic gravitational wave background which would also be a test of strong gravity for distinguishing sources \cite{Kirillov:2021bcs}. 
	
	In this article, we attempt to constrain few well-known WH solutions as a test for strong gravity. That is, by choosing specific WHs which are exact solutions of certain gravitational theories, the free parameters such as charge and throat radius, for each WH spacetime are constrained using the data of S2 star orbiting Sgr A$^\star$. We also reconstruct shadow images for each WH and make comparison with a shadow of Schwarzschild BH since we work with static and spherical symmetric spacetimes only. Our paper is structured as follows. In Sec.~\ref{secorb} we give the overview of different WH geometries in consideration and analyze their embedding diagrams in a flat Euclidean space. In Sec.~\ref{OD}, we will prescribe the procedure to study orbital dynamics of S2 star in the WH geometries. In Sec.~\ref{seccons} we discuss and determine the constraints on the parameters of the three WH models introduced in our previous sections. In Sec.~\ref{secacc} we consider the accretion of infalling gas and study the mass accretion rate and how it is related to the throat rate change. We evaluate the intensity of the electromagnetic radiation due to the accretion of thin disk and compare it to the energy radiated by the same disk accreting onto a Schwarzschild BH of same mass as the WH. In Sec.~\ref{secshad} we investigate the shadow images and the intensity of WHs using infalling gas model. In Sec.~\ref{secapp} we firstly produce synthetic datasets and then we reconstruct the images of WHs using the infalling gas model. We conclude in Sec.~\ref{secconc}.
	
\section{WH models}
\label{secorb}
	
Let us consider a static and spherically symmetric spacetime ansatz, commonly termed as Morris-Thorne traversable WH, which in Schwarzschild coordinates can be written as follows~\cite{Morris:1988cz}
\begin{equation}
			ds^2=A(r)dt^2-\frac{dr^2}{B(r)}-C(r)\left(d\theta^2+\sin^2\theta \,d\phi^2\right)
\end{equation}
where 
\begin{equation*}
A(r)=e^{2\Phi(r)},~~B(r)=1-\frac{b(r)}{r},~~C(r)=r^2.
\end{equation*}%
Here $\Phi(r)$ is the redshift function  and $b(r)$ is the shape function. These functions depend on the radial coordinate $r$ and on the parameters of the solution as the throat radius $r_0$. In subsequent sections we will need to write explicitly, say, $b(r,r_0)$ and in this case a prime notation will always mean derivative with respect to $r$ (this derivative is sometimes written as $\partial /\partial r$) and derivative with respect to $r_0$ will always be denoted by $\partial /\partial r_0$.  The throat corresponds to the minimum value of $r^2$. The redshift function $\Phi (r)$ should be finite in order to avoid the formation of an event horizon and should tend to zero for large $r$ to ensure asymptotic flatness. On the other hand the shape function $b(r)$ determines the WH geometry, with the following condition $b(r_{0})=r_{0}$. Consequently, it follows that the functions ($b,\,e^{2\Phi(r)}$) have to satisfy the following conditions:
\begin{align}
		&\lim_{r\to\infty}e^{2\Phi(r)}=\text{finite}=1,\nonumber\\
		&b<r\text{ if }r>r_0\;\text{ and }\;b(r_0)=r_0,\nonumber\\
		\label{b2}&\lim_{r\to\infty}(b/r)=0,\\
		&rb'<b\;(\text{flaring-out condition}),\nonumber\\
		&b'(r_0)\leq 1.\nonumber
\end{align}
	If the mass $\mathcal{M}$ of the WH is finite then one must have $\lim_{r\to\infty}b=2\mathcal{M}$~\cite{Visser:1995}.

	\subsection{Model I: Specific redshift and shape functions}
	Let us consider the following choice for the WH metric \cite{Shaikh:2018kfv,Gyulchev:2018fmd,Jamil:2009vn}
	\begin{equation}
		ds^2=e^{-2\left( \frac{r_0}{r}+\frac{r_0^2}{r^2} \right)}dt^2-\frac{dr^2}{1-\frac{b(r)}{r}}-r^2\left(d\theta^2+\sin^2\theta \,d\phi^2\right).
	\end{equation}
	where the redshift function is given by
	\begin{eqnarray}
		\Phi(r)=-r_0/r-r_0^2/r^2,
	\end{eqnarray}
	and 
	\begin{equation}
		b(r)=r_0 \left( \frac{r_0}{r} \right)^{\gamma},
	\end{equation}
	where $\gamma\geq 0$. If $\gamma =0$, the mass $\mathcal{M}=r_0/2$ and in the case $\gamma > 0$ we have $\lim_{r\to\infty}b=0=2\mathcal{M}$, so that the WH is massless.
	If $\gamma <0$, the WH mass can diverge, i.e. $\mathcal{M}=\infty$. Accordingly, one can chose only finite distance corrections and in general the case $\gamma <0$ is a mathematical curiosity and not a physical solution. Of course, for such a solution one can always adjust the parameters to get the desired best fit.
	
	Let us recall that the rotational velocity of a test particle in spherically symmetric space-time, within the equatorial
	plane is determined by 
	\begin{eqnarray}\label{vtg}
		v_{tg}^2(r)=r \,\Phi'(r),
	\end{eqnarray} 
	yielding
	\begin{eqnarray}
		v_{tg}^2(r)=\frac{r_0}{r}\left[1+2\left(\frac{r_0}{r}\right) \right].
	\end{eqnarray}
	As a special case $\gamma=0$ and $\Phi(r)=-r_0/r$, we obtain a special case having $v_{tg}^2(r)=\frac{r_0}{r}$, which was studied by Bambi \cite{Bambi:2013nla}. 
	
	\subsection{Model II: The Bornnikov-Kim WH solution}
	An exact  WH solution in the context of Einstein-Dirac-Maxwell theory has been recently proposed in \cite{Blazquez-Salcedo:2020czn} which coincides with the Bronnikov-Kim WH solution obtained as an exact solution in the context of brane-world gravity \cite{Bronnikov:2002rn,Bronnikov:2003gx} given by
	\begin{equation}\label{Dirac}
		ds^2=\left(1-\frac{M}{r}\right)^2 dt^2-\frac{dr^2}{\left(1-\frac{r_0}{r}  \right) \left(1-\frac{Q^2}{r_0 r}   \right) }-r^2d\Omega^2,
	\end{equation}
	where $d\Omega^2=d\theta^2+\sin^2\theta \,d\phi^2$ and the parameter $M$ is related to $Q^2$ by
	\begin{eqnarray}
		M=\frac{ 2Q^2 r_0}{Q^2+r_0^2},
	\end{eqnarray}
	Note that $r_0$ denotes the size of WH throat (since one assumes $r_0>Q$) and $Q$ is the charge. As we pointed out, this metric coincides with Bronnikov-Kim WH upon making the substitutions $M\to 2M$ and $Q^2/r_0\to r_1$, yielding \cite{Bronnikov:2002rn,Bronnikov:2003gx}
	\begin{equation}\label{whm2}
		ds^2=\left(1-\frac{2M}{r}\right)^2 dt^2-\frac{dr^2}{\left(1-\frac{r_0}{r}  \right) \left(1-\frac{r_1}{r}   \right) }-r^2d\Omega^2,
	\end{equation}
	where $r_0$ is the throat's radius [$r_0>r_1\equiv Mr_0/(r_0-M)$]. If we now introduce the following dimensionless parameter $q=r_0/2M>1$ \cite{Bronnikov:2021liv}, then the condition for having a WH is provided by $q>1$.

	Note that for the solution~\eqref{Dirac} the effective shape function has the form
	\begin{eqnarray}
		b(r)=r_0+\frac{Q^2}{r_0}-\frac{Q^2}{r},
	\end{eqnarray}
	and the mass $\mathcal{M}$ of the WH (\ref{whm2}) is given by
	\begin{equation}\label{mass}
		\mathcal{M}=\frac{Q^2+r_0^2}{2r_0}.	
	\end{equation}
	\subsection{Model III: Braneworld gravity WH}
	Another interesting  WH solution in brane-world gravity found in \cite{Bronnikov:2002rn,Bronnikov:2003gx} is given by (also known as the Casadio-Fabbri-Mazzacurati metric)
	\begin{equation}\label{WHIIIa}
		ds^2=\left(1-\frac{2r_g}{r}\right) dt^2-\frac{1-\frac{3r_g}{2r} }{\big( 1-\frac{r_0}{r} \big)\left(1-\frac{2r_g}{r}\right)}dr^2-r^2d\Omega^2,
	\end{equation}
	where $r_0>2r_g$ and $r_g$ is defined by
	\begin{eqnarray}\label{WHIIIb}
		r_g\equiv\frac{MG}{c^2},
	\end{eqnarray}
	where $G$ and $c$ are known universal constants. The shape function has the form
	\begin{eqnarray}\label{WHIIIc}
		b(r)=r-\frac{\big( 1-\frac{r_0}{r} \big)\left(1-\frac{2r_g}{r}\right)}{1-\frac{3r_g}{2r}}r\quad\text{and}\quad r_0>2r_g,
	\end{eqnarray}
	yielding the following relation between the parameter $M$ and mass of the WH is
	\begin{equation}\label{mass2}
		\mathcal{M}=\frac{c^2r_0}{2G}+\frac{M}{4}=\frac{c^2(2r_0+r_g)}{4G}.	
	\end{equation}
	
	Note that metric \eqref{WHIIIa} can describe different objects depending on the value of $r_0$. In particular if we introduce the following dimensionless parameter $q=r_0/r_g>0$ \cite{Bronnikov:2021liv}, one can have a traversable WH if $q \in (2, \infty)$ or a regular BH if $q \in (\frac{3}{2}, 2)$. The Schwarzschid BH is recovered if $q=3/2$ yielding $\mathcal{M}=M$, and the case $q \in (0, \frac{3}{2})$ corresponds to a Schwarzschild-like BH with the curvature singularity lying at $r=\frac{3}{2}r_g$.
	
	\begin{figure*}
		\includegraphics[width=0.32\textwidth]{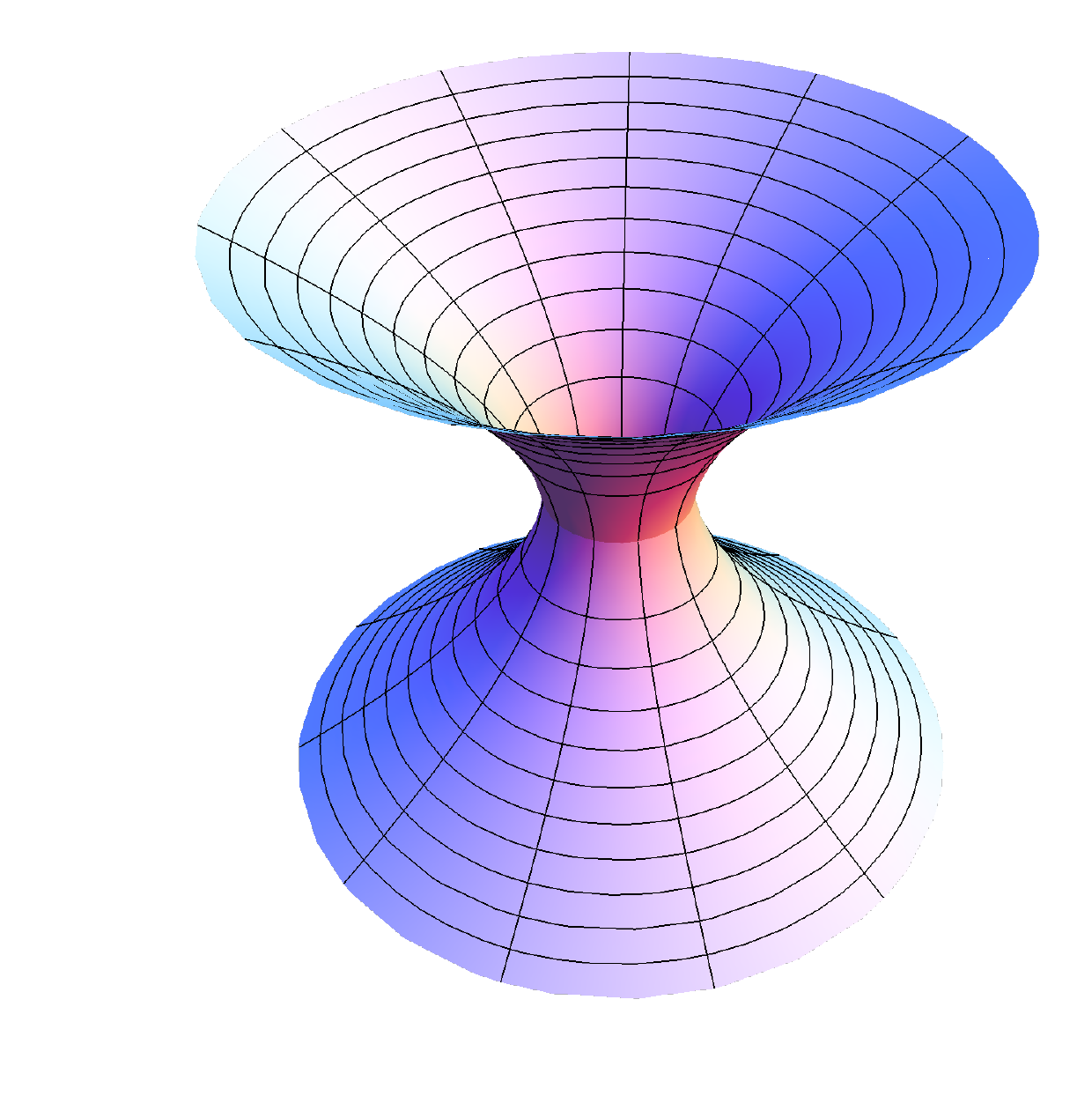}
		\includegraphics[width=0.32\textwidth]{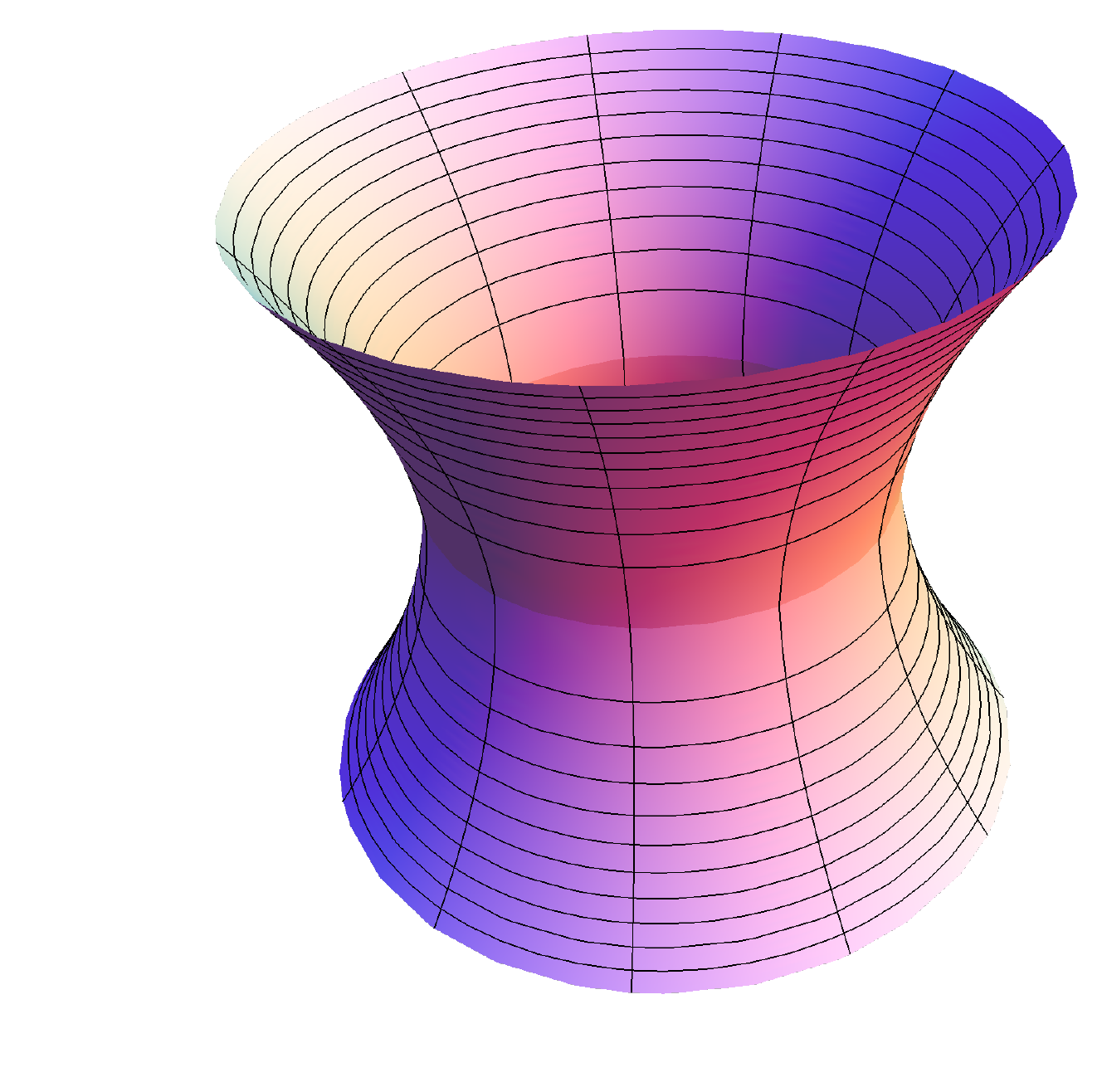}
		\includegraphics[width=0.32\textwidth]{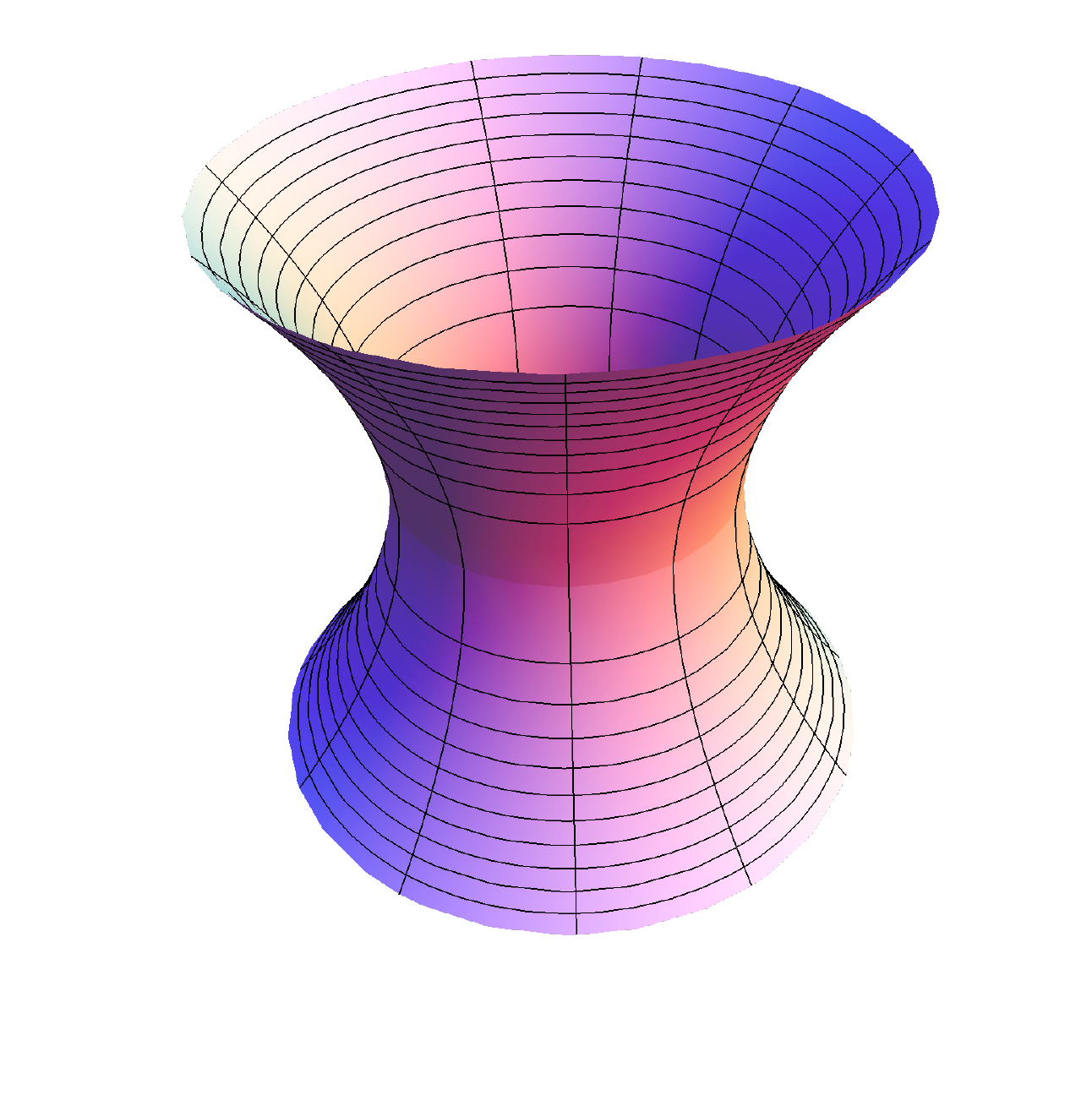}
		\caption{Left to right: The embedding diagram of WH model I, WH model II and WH model III, respectively. We have used the constraint values for parameters obtained from S2.}
	\end{figure*}
	
	For a physical definition of the mass of an attractive center, here a wormhole, we have adopted the ADM one~\cite{Visser:1995}, which is the commonly used definition and it is more relevant for theoretical investigations. From an observational point of view, the ADM mass can be related to many other physical entities that can be directly observed (as the coefficient of the $1/r$ term in the expansion of $g_{tt}$ or the coefficient of the $1/r^{3/2}$ term in the orbital frequency).

	\subsection{Embedding of WH slice in flat space}
	
	Having the expression for $b(r)$ one can discuss the embedding diagrams to represent the above WH models by considering an equatorial slice $\theta=\pi/2$ at some fix moment in time $t=$ constant. The metric can be written as 
	\begin{equation}
		ds^2=\frac{dr^2}{1-\frac{b(r)}{r}}+r^2d\phi^2.\label{emb}
	\end{equation}
	We embed the metric (\ref{emb}) into three-dimensional Euclidean space (written in cylindrical coordinates) to visualize this two dimensional spatial slice  
	\begin{equation}
		ds^2=dz^2+dr^2+r^2d\phi^2.
	\end{equation}
	From the last two equations we find that 
	\begin{equation}
		\frac{dz}{dr}=\pm \sqrt{\frac{r}{r-b(r)}-1}.
	\end{equation}
	where $b(r)$ depends on the specific choice of the WH geometry. Note that the integration of the last expression cannot
	be accomplished analytically. Invoking numerical techniques allows us to illustrate the
	WH shape given in Fig. (1). 
	
	\section{Orbital dynamics}\label{OD}
	The equations of motion of the test particle (S2 star), in the spherically symmetric WH  metric assuming without loss of generality $\theta= \pi/2$, are: 
		\begin{subequations}
			\begin{eqnarray}
				\dot{t} &=& \dfrac{\mathcal{E}}{A(r)},\label{eqn:motiont}\\
				\ddot{r} &=& \frac{-B(r)}{2} \left[A'(r) \ \dot{t}^2 + \frac{B'(r)}{B(r)^2} \ \dot{r}^2 - C'(r)  \dot{\phi}^2\right],\label{eqn:motionr}\\
				\dot{\phi} &=&  \dfrac{\mathcal{L}}{r^2},\label{eqn:motionphi}
			\end{eqnarray}
	\end{subequations}
	where $\mathcal{E}$ and $\mathcal{L}$ are the conserved energy and the angular momentum of the particle per-unit-mass, and the overdot stands for derivative with respect to the proper time, $\tau$. 
	In terms of Cartesian coordinates, we denote the position of the real orbit as $(x, y, z)$, and velocity components $(v_x, v_y , v_z)$. In our present case, $\theta = \pi/2$, these are obtained using the transformation from spherical Schwarzschild coordinates to Cartesian coordinates:
	\begin{equation}\label{eqn:xyz}
		x = r \cos\phi ,\qquad y = r \sin\phi ,\qquad z = 0,
	\end{equation}
	and the corresponding three-velocities are \cite{Becerra-Vergara:2020xoj}:
	\begin{equation}\label{eqn:vxvyvz}
		v_x = v_r \cos\phi - r v_\phi \sin\phi,\,\,v_y = v_r \sin\phi + r v_\phi \cos\phi,\,\, v_z = 0,
	\end{equation}
	where $v_r= dr/dt$ and $v_\phi= d\phi/dt$.

	One can use eqs.~(\ref{eqn:motiont})--(\ref{eqn:motionphi}) to obtain the orbit of S2 star and then we can compare with the observational data to constrain the parameters in our WH models. To do so, we also need to find the apparent orbit on the plane of the sky by projecting the real orbit onto the observation plane as was argued in ~\cite{Becerra-Vergara:2020xoj}. {On the plane of the sky, the star traces an orbit with Cartesian positions $X_{\rm obs}$ and $Y_{\rm obs}$, defined by the observed angular positions, i.e. the declination $\delta$ and the right ascension $\alpha$ 
		\begin{equation}\label{eqn:XY}
			X_{\rm obs} = D_\odot (\alpha-\alpha_{\rm Sgr A*}),\quad Y_{\rm obs} = D_\odot (\delta-\delta_{\rm Sgr A*})
		\end{equation}
		centering the coordinate system on Sgr A$^\star$. We adopt in this work $D_\odot = 8$~kpc
	}

	Following the same arguments shown in \cite{Becerra-Vergara:2020xoj} and without going into further details here we can relate the apparent orbit given by coordinates $(\mathcal{X}, \mathcal{Y}, \mathcal{Z})$ to the real orbit given by $(x, y, z)$ and obtain 
	the corresponding components of the apparent coordinate velocity, $(\mathcal{V}_X=d\mathcal{X}/dt, \mathcal{V}_Y=d\mathcal{Y}/dt, \mathcal{V}_Z=d\mathcal{Z}/dt)$,  as follows  \cite{Becerra-Vergara:2020xoj}
	\begin{eqnarray}\notag
		\mathcal{V}_X &=& v_x \left(\sin\Omega \cos\omega + \cos\Omega \sin\omega \cos i\right)\\\notag
		&+& v_y\left(-\sin\Omega \sin\omega + \cos\Omega \cos\omega \cos i\right),\\\notag
		\mathcal{V}_Y &=& v_x \left(\cos\Omega \cos\omega - \sin\Omega \sin\omega \cos i\right)\\\notag
		&+& v_y \left(-\cos\Omega \sin\omega - \sin\Omega \cos\omega \cos i\right),\\
		\mathcal{V}_Z &=& v_x \left(\sin\omega \sin i\right) + v_y \left(\cos\omega \sin i \right),
	\end{eqnarray}
	in which we have the following quantities: $\omega$, $i$, and $\Omega$ (the \emph{osculating} orbital elements) known as the argument of pericenter, the inclination between the real orbit and the observation plane, and the ascending node angle, respectively. It's worth noting that the radial position can be written as 
	\begin{equation}
		r = a (1-e \cos E),
	\end{equation}
	with $a$ being the semi-major axis of the ellipse, $e$ the eccentricity, and $E$ its eccentric anomaly. The latter is related to the true anomaly, which is the azimuthal angle $\phi$, by
	\begin{equation}
		\cos\phi = \frac{(\cos E - e)}{(1-e\cos E)}.
	\end{equation}
	Unfortunately, as we know in the case of general relativistic setup, finding an exact analytic expression in a closed-form for $r(\phi)$ is not possible. In order to compare and test different theories we can use  $r(t)$ and $\phi(t)$ via a numerical integration of the equations of motion~\eqref{eqn:motiont}. That means we can find $r(\phi)$ by means of numerical methods.
	\begin{figure}
		\centering
		\includegraphics[width=2.8 in]{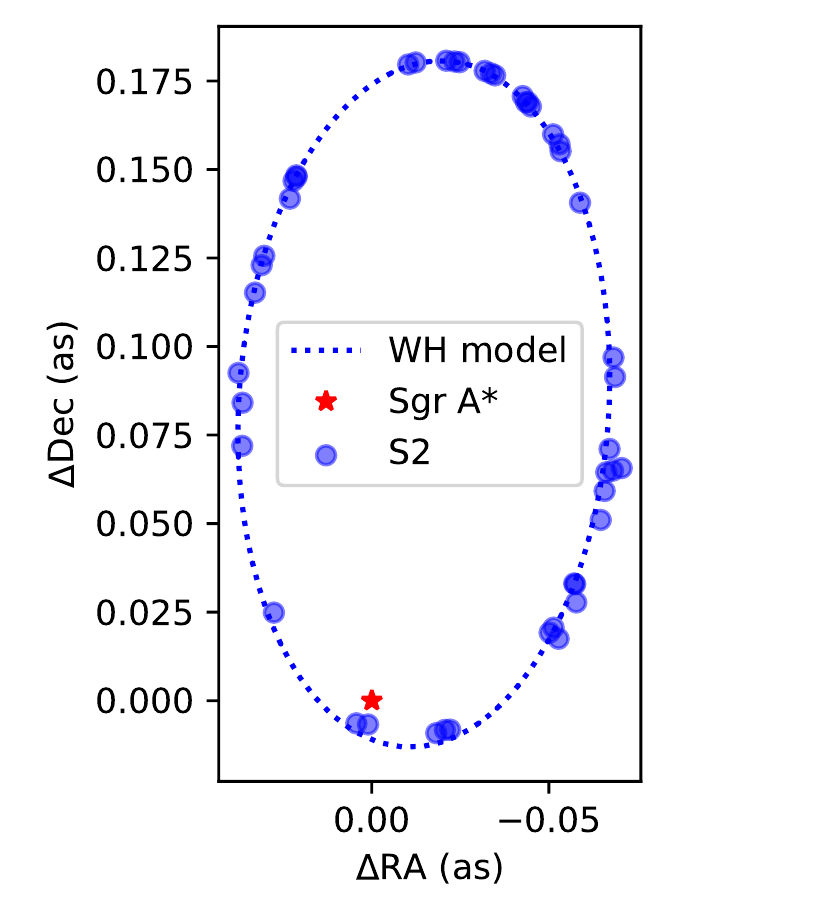}
		\caption{Orbit of S2 around the galactic center (Sgr A$^\star$) as a model fitting using the WH model I. We used the observational data from \cite{Do:2019}. In the particular case we used the best fit $r_0=1.03 M$ and $\gamma=-0.24$. We work in mass units $4.07 \times 10^6 M_{\odot}=1$.} 
		\label{orbit1}
	\end{figure}
	
	\section{Constraints using S2 star orbit}
	\label{seccons}
	In this section, we shall turn our attention to study the WH geometries using observational data for the star S2. In fact, one of the methods to study the nearby geometry of the central compact object in the Milky Way galactic center is to analyze orbits of S-stellar cluster around Sgr A$^{\star}$ \cite{nucita,nucita1,zaka,maria,babichev}. In fact, it was argued that one can use the S cluster stars to set constraints on the BH mass (in the present work we shall assume units of mass $4.07\times 10^6 M_{\odot}$ \cite{Gillessen:2017}).  Moreover the motion of S2 star has been used to constrain different models for the dark matter distribution inside the inner galactic region, such as the dark matter spike model investigated in Ref. \cite{Nampalliwar:2021tyz}. To this end, we are going to use the S2 orbit data collected during the last few decades (see, \cite{Do:2019}), to fit the different WH models, in particular we would like to see to what extent the WH models considered here can mimic the BH hole geometry. 
	
	We shall analyze the motion of S2 star using the WH geometry. To fit our WH model, we have to solve the equations of motion numerically (see, for details  \cite{rueda}) and to use few orbit parameters, including the inclination angle ($i$), argument of periapsis  ($\omega$), angle to the ascending node ($\Omega$), the semi-major axis ($a$) and eccentricity ($e$) of the orbit. The best-fitting values for the parameter $b$ and the BH mass $M$ are derived from the MCMC analysis using the emcee software. Note that in our setup, we use the Bayesian theorem according to which the observations O, and the vector containing the parameters of a model, say P, the posterior probability density $\pi (P | O)$, is given by  \cite{Grould:2017bsw}
	\begin{eqnarray}
		\ln \pi(P|O) \propto \ln f(O|P)+\ln \pi(P),
	\end{eqnarray}
	in which $ \pi(P)$ is the prior probability density of the parameters and  the likelihood function is given by 
	\begin{eqnarray}
		\ln f(O|P)=-\frac{1}{2}\sum_{i=1}^{N}\left[ \frac{\left(\alpha_{obs,i}-\alpha_{mod,i}\right)^2}{\sigma_{obs,i}^2} \right],
	\end{eqnarray}
	where $\alpha_{obs}$ and $\alpha_{mod}$ are the two observed and theoretical(WH models) quantities $(X_{obs},Y_{obs})$, and $(X_{mod},Y_{mod})$, respectively. In what follows we shall present our results for the each WH model.
	\subsection{WH model I}
\begin{figure*}
		\includegraphics[width=0.33\textwidth]{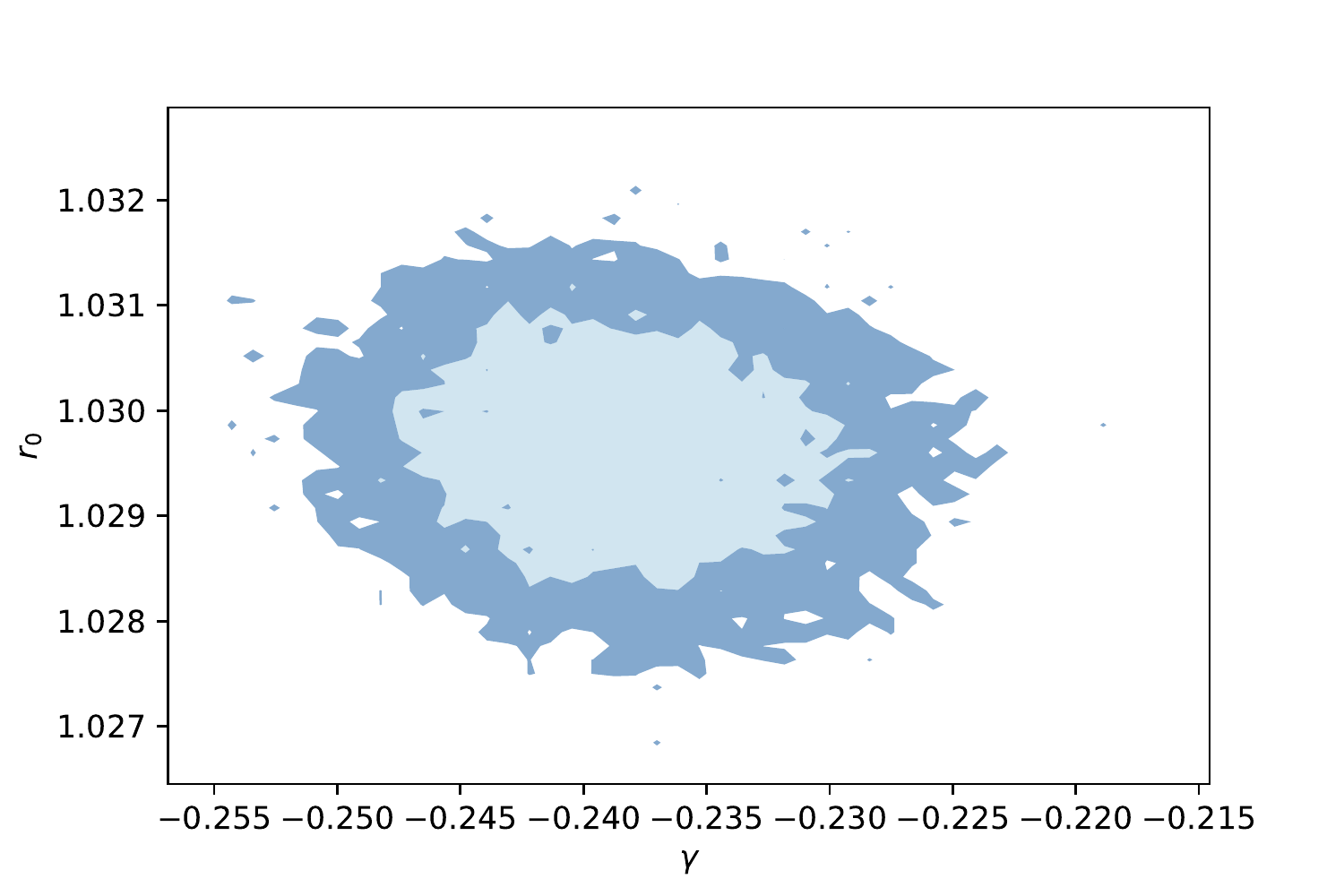}
		\includegraphics[width=0.33\textwidth]{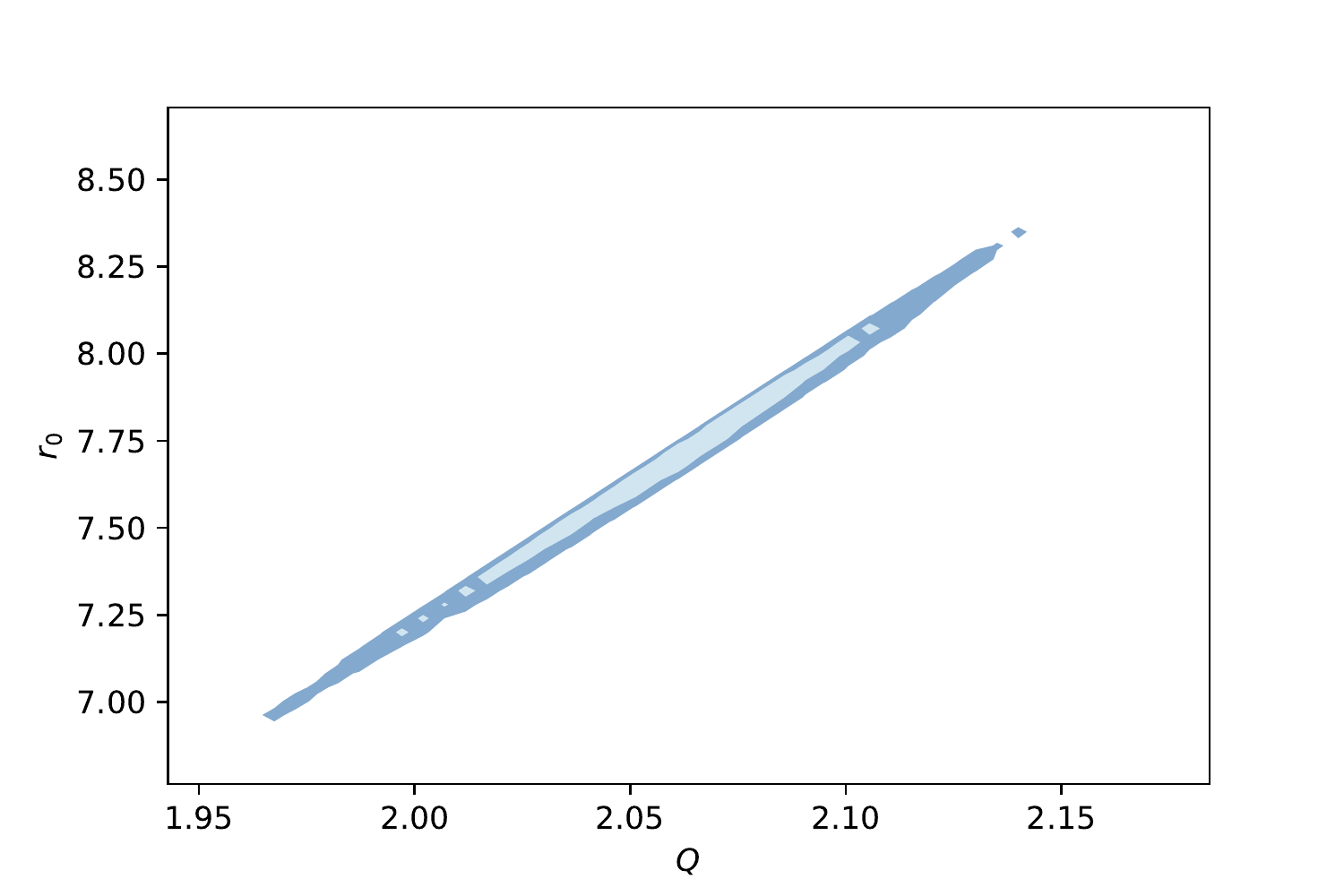}
		\includegraphics[width=0.33\textwidth]{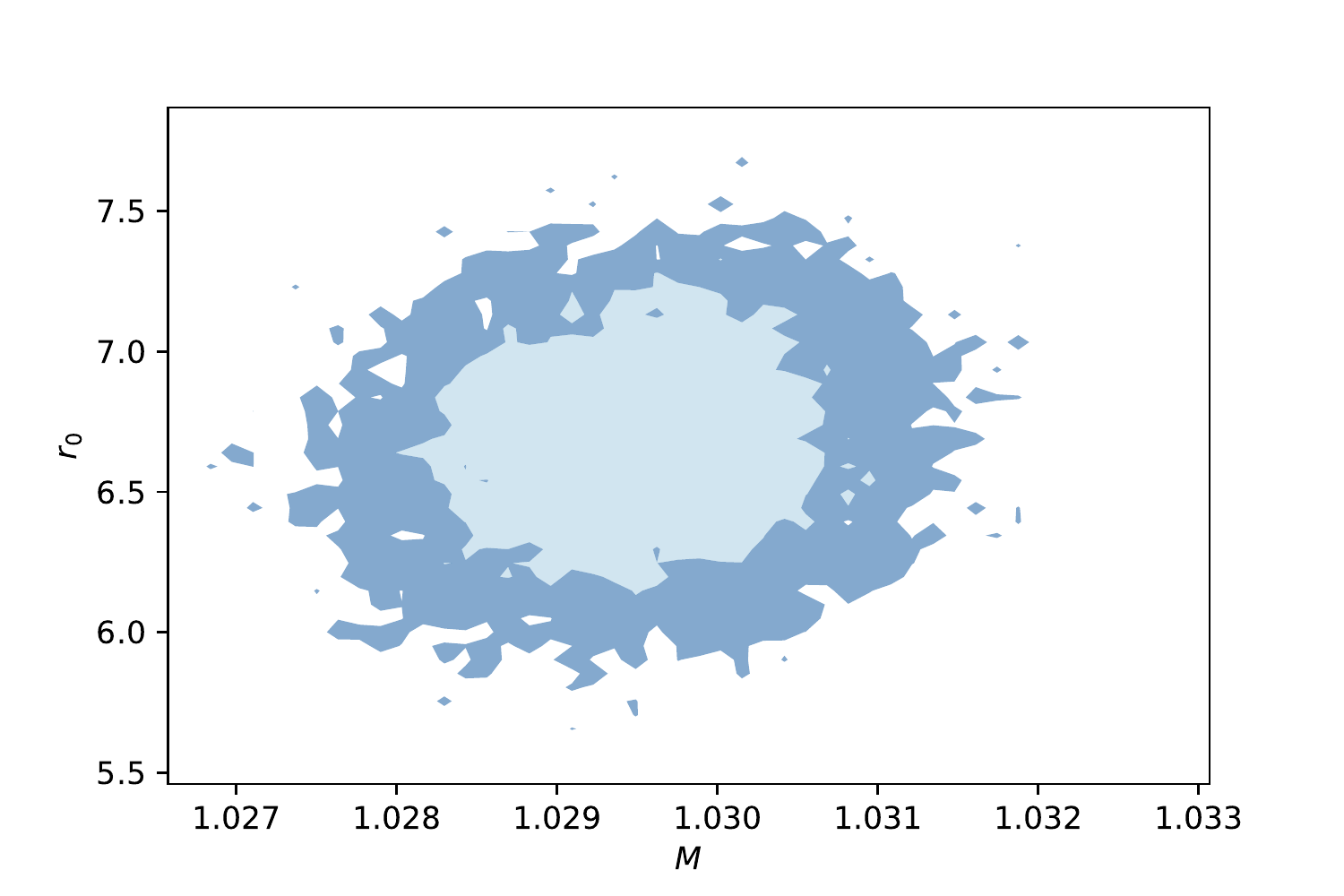}
		\caption{The constrains of WH parameters $(r_0,\gamma, Q, M)$ using the parameter space for three different WH models with 68\% and 96\% confidence contours.}\label{contourwh}
	\end{figure*}

	Using the WH model I along with the observational data we obtain numerically the best-fitting orbit for the S2 as shown in Fig.~\ref{orbit1}, where the star $\star$ denotes the position of Sgr A$^\star$ or the galactic center. In this specific model, we took the uniform priors within the range $r_0 \sim [M,3 M]$ and $\gamma \sim [-1,1] $ and found the best fitting values for WH model I  within $68\%$ confidence level $r_0=1.029^{+0.003}_{-0.001} M$ as shown in Fig. \ref{contourwh} (a). We obtain for the parameter $\gamma$ the best fitting value $\gamma=-0.238^{+0.006}_{-0.005}$ within $68\%$ confidence level. In order to maintain the WH structure, we need to check whether the flaring out condition is satisfied, that is such a condition needs to be satisfied in order to keep the WH mouth open. This condition at the WH throat region is given by the following relation 
	\begin{align}
		b'(r_0)=-\frac{\gamma r_0}{r} \left(\frac{r_0}{r}\right)^{\gamma}\Big|_{r_0}<1.
	\end{align}
	
	Taking $r=r_0\sim 1.03$  for the WH throat radius and $\gamma \sim -0.24$, we find that this condition is indeed satisfied for this model as well
	\begin{align}
		b'|_{r_0}\sim 0.24<1.
	\end{align}
	
	Note also that for the eccentricity  and the  semi-major axis we obtain $0.89$ and $126.4$ mas, respectively. 
	
	\subsection{WH model II}
	For the WH throat we found the best fitting values within $68\%$ confidence level $r_0=7.71^{+0.28}_{-0.40} M$ as shown in Fig. \ref{contourwh} (b). Moreover, we also find from the parameter space the best fitting values for the charge  $Q=2.06^{+0.04}_{-0.05}$ within $68\%$ confidence level. We find that this condition is indeed satisfied for this WH model
	\begin{eqnarray}
		b'(r_0)=\frac{Q^2}{r^2}\Big|_{r_0},
	\end{eqnarray}
	yielding
	\begin{align}
		b'|_{r_0}\sim 0.07<1.
	\end{align}
	
	At this point, it is interesting  to estimate here also the mass parameter using the best fit values for $(Q,r_0)$
	\begin{eqnarray}
		M=\frac{ 2Q^2 r_0}{Q^2+r_0^2} \sim 1.028
	\end{eqnarray}
	implying that the condition $Q>M$ is fulfilled. Alternatively, in the language of the metric \eqref{whm2} we can estimate $r_1\sim 0.55$ implying the following conditions: $r_0>r_1$ and $q=r_0/2M\sim 3.74 >1$, are also satisfied.
	
	\subsection{WH model III}
	We found the best fitting values within $68\%$ confidence level $r_0=6.639^{+0.394}_{-0.295} M$ as shown in Fig. \ref{contourwh} (c). For the parameter space we obtain the best fitting values $M=1.029^{+0.001}_{-0.001}$ within $68\%$ confidence level. We find that this condition is indeed satisfied for this model too
	\begin{eqnarray}
		b'(r)=\frac{M\left(2 r_0-3M\right)}{(3M-2r_0)^2}\bigg|_{r_0}
	\end{eqnarray}
	at the WH throat yielding
	\begin{align}
		b'|_{r_0}=0.100<1.
	\end{align}
	
	Thus, we can estimate the ratio between $r_0$ and $M$ to obtain the quantity $q=r_0/M \sim 6.451$. In other words, for such domain of parameters the metric \eqref{WHIIIa} describes a traversable WH spacetime.
	
	Finally we have used the reduced $\chi^2$ to see how well the WH models fit the data using the following definition \cite{Grould:2017bsw}
	\begin{eqnarray}
		\chi^2=\sum_{i=1}^{N}\left[ \frac{\left(\alpha_{obs,i}-\alpha_{mod,i}\right)^2}{\sigma_{obs,i}^2} \right],
	\end{eqnarray}
	along with
	\begin{eqnarray}
		\chi_r^2=\frac{\chi^2}{k},
	\end{eqnarray}
	where $k=3N-n$, with $N$ being the number of observation dates and $n$ the number of fitted parameters. Sometimes it is useful to compute the mean of the $\chi_r^2$ defined as $<\chi^2>=(\chi_r^2)/2=(\chi^2_X+\chi^2_Y)/2$. In Table I we present our results which suggest that in all models $\chi_r^2>1$, while the best fitting model is achieved when $\chi_r$ is around 1. That means WH model I and WH model III fits the data well and are better models compared to WH model II and the BH model.  In other words, based on the motion of S2 orbit WH geometry can mimic the BH geometry very well but we can distinguish them via the shadow images. In what follows we shall comment on accretion of matter onto WHs using the best fits value we obtained above.

	\begin{table*}
		\begin{center}
			\begin{tabular}{|l|l|l|l|l|l|l|}
				\hline
				&\,\,\,\,WH model I &\,\,\,\,WH model II  &\,\,\,\,WH model III &\,\,\,\,BH \\ 
				\hline
				Eccentricity\,\,\,[no units] & 0.890  &  0.890  & 0.890     &  0.890  \\ 
				Inclination\,\,[$^{0}$] & 134.567 & 134.567   &  134.567     & 134.567  \\ 
				Ascending Node\,\,[$^{0}$] & 228.171  &  228.171  &  228.171    & 228.171  \\ 
				Semimajor Axis\,\, [as] & 0.1264  &   0.1264   &  0.1264     &   0.1264  \\
				Wormhole throat\,\,[M]  & 1.03 &   7.70   &  6.64     &   /  \\
				Mass parameter $[ 4.07 \times 10^6 M_\odot]$   &  / &    1.028  &    1.03   &    1.03 \\
				Charge parameter\,\,[M]  &  / &   2.06  &    /   & /  \\
				$\gamma$ parameter\,\,[no units]  &  -0.24 &   /  &    /   & /  \\
				Shadow radius\,\,[M]  &  4.36 &   8.88  &    7.99   & 5.19  \\
				\,\,\,\,\,\,\,\,\,\,$\chi_r^2$ & 5.89 & 6.25 &  5.97    & 7.47\\
				\,\,\,\,$<\chi^2>$ & 2.94 & 3.12 &  2.98    & 3.73\\
				\hline
			\end{tabular}
			\caption{Values of different parameters used and $\chi_r^2$ for each WH model and the black hole, respectively.}
		\end{center}
	\end{table*}

	\section{Accretion of matter onto WH}
	\label{secacc}
	In this section we first present a general treatment by which we relate the mass accretion rate $\dot{\mathcal{M}}$ (with $\mathcal{M}$ being the mass of the WH and dot denotes time derivative), of any accretion process spherical or not, to the rate of change of the throat radius $\dot{r}_0$.
	
	We take the stress-energy tensor (SET) of a WH to be anisotropic of the form 
	\begin{equation}\label{a0}
		T^{\mu}{}_{\nu}={\rm diag}(\rho(r),-p_r(r),-p_t(r),-p_t(r)),
	\end{equation}
	with no dissipation effects, where $\rho$ is the energy density and $p_r$ and $p_t$ are the radial and transverse pressures respectively. Using solely the Einstein field equations and ($G^{\mu}{}_{\nu}=8\pi T^{\mu}{}_{\nu}$), the SET conservation
	\begin{align}\label{a1}
		&\frac{\partial b}{\partial r}=8\pi r^2\rho ,\\
		&2\frac{\partial \Phi}{\partial r}=\frac{8 \pi  r^3p_r +b}{r (r-b)},\\
		&2 p_t=2 p_r+ r \frac{\partial p_r}{\partial r}+r(p_r+\rho )\frac{\partial \Phi}{\partial r},
	\end{align}
	and the relations $b(r_0)=r_0$ and $\lim_{r \to \infty}b=2\mathcal{M}$, it is straightforward to show that~\cite{Azreg:2015,Visser:1995}
	\begin{equation}\label{a2}
		2\mathcal{M}=r_0 + 8\pi\int_{r_0}^{\infty}r^2\rho(r)\,d r .
	\end{equation}
	In the case when $\rho\geq 0$ for all $r$, we obtain $\mathcal{M}\geq r_0/2$~\cite{Azreg:2015}. This static relation holds also in any accretion process (spherical or not) provided the accreting matter [the interstellar gas need not to be confused with the source term in the field equations~\eqref{a0}] does not alter the geometry of the spacetime nor does it alter the SET~\eqref{a0}. That is, the accretion process is slow enough to permit the spherical symmetry of the WH and the SET remains preserved. However, if $\mathcal{M}\geq r_0/2$ holds during an accretion process, this does not mean that $\dot{\mathcal{M}}$ and $\dot{r}_0$ have necessarily the same sign. In the case where $\rho$ may assume both signs, any of the two static relations $\mathcal{M}\gtrless r_0/2$ holds during an accretion process. To show that this is indeed the case, we will shortly derive the relation $\mathcal{M}(t)/r_0(t)=\mathcal{M}/r_0$ where ($\mathcal{M},\,r_0$) are some initial values.
	
	During an accretion process the mass $\mathcal{M}$ becomes a slightly varying function of time where the accretion rate $\dot{\mathcal{M}}$ may have both signs depending on the nature of the accreting matter (ordinary matter, dark matter or phantom matter). To see how the throat radius varies during an accretion process, we differentiate both sides of~\eqref{a2} with respect to time 
	\begin{equation}\label{a4}
		2\dot{\mathcal{M}}=\Big(1-8\pi r_0^2\rho_0 + 8\pi\int_{r_0}^{\infty}r^2\frac{\partial\rho(r,r_0)}{\partial r_0}\,d r \Big)\dot{r}_0 , 
	\end{equation}
	where we assumed that the energy density, being a function of the radial coordinate $r$, depends also on $r_0$: $\rho\equiv\rho(r,r_0)$. This is obvious from the first line in~\eqref{a1} since the function $b$ always depends on $r_0$. 
	By the first line in~\eqref{a1} we have $\partial^2 b/\partial r_0\partial r = 8\pi r^2 \partial\rho(r,r_0)/\partial r_0$. Using this in~\eqref{a4} and integrating we obtain
	\begin{eqnarray}\label{a5}\notag
		2\dot{\mathcal{M}}&=&\bigg(1-8\pi r_0^2\rho_0 + \frac{\partial b}{\partial r_0}\bigg|_{r=r_0}^{r\to\infty} \,\bigg)\dot{r}_0\\
		&=&\bigg(1-\frac{\partial b}{\partial r}\bigg|_{r=r_0} + \frac{\partial b}{\partial r_0}\bigg|_{r=r_0}^{r\to\infty} \,\bigg)\dot{r}_0 ,
	\end{eqnarray}
	where we have used~\eqref{a1}.
	
	In order to evaluate the right-hand side of~\eqref{a5} we need to notice that the general expression of $b(r,r_0)$ may be brought to the form
	\begin{equation}\label{a6}
		b(r,r_0) = r_0 h(y),\quad h(0)=\frac{2\mathcal{M}}{r_0},\quad h(1)=1\qquad \Big(y\equiv \frac{r_0}{r}\Big),
	\end{equation}
	provided the mass $\mathcal{M}$ of the WH is finite. We obtain
	\begin{equation}\label{a7}
		\frac{\partial b(r,r_0)}{\partial r}\bigg|_{r=r_0}=-h'(1),
	\end{equation}
	(here $h'\equiv dh/dy$) and
	\begin{equation}\label{a8}
		\frac{\partial b}{\partial r_0}=h(y)+yh'(y).
	\end{equation}
	This yields \[\frac{\partial b}{\partial r_0}\Big|_{r\to\infty}=h(0)=\frac{2\mathcal{M}}{r_0}\qquad\text{and}\qquad \frac{\partial b}{\partial r_0}\Big|_{r=r_0}=h(1)+h'(1),\] where we assumed that $b$ is not perturbed during accretion since the mass of the accreting matter is supposed to be much smaller than that of the attractive center. Substituting in~\eqref{a5} we arrive at
	\begin{equation}\label{a9}
		\dot{\mathcal{M}}	=\frac{\mathcal{M}}{r_0}~\dot{r}_0.
	\end{equation}
	We conclude that $\dot{\mathcal{M}}$ and $\dot{r}_0$ have the same sign provided $M>0$; for WHs with negative mass this conclusion no longer holds and $\dot{\mathcal{M}}$ and $\dot{r}_0$ evolve in opposite directions. By integration we obtain
	\begin{equation}\label{a10}
		\frac{\mathcal{M}(t)}{r_0(t)}=\frac{\mathcal{M}}{r_0},
	\end{equation}
	where ($\mathcal{M},\,r_0$) denote the initial values prior to accretion.
	
	In the forthcoming part of this section we aim to evaluate the mass function $\mathcal{M}(t)$ taking as models the well-known spherical accretion and thin accretion disk. 
	
	\subsection{Spherical accretion}
	If the accreting matter forms a large gas cloud from the interstellar medium and the central object (star, WH or BH) is isolated, then spherical accretion is a reasonable approximation to the real situation. Spherical accretion in its general case onto a central object the geometry of which is described by the metric Eq. 1,
	has been treated in~\cite{Azreg:2018}. It was shown that the mass accretion rate $\dot{\mathcal{M}}$ is proportional to the value $h_\infty$ of the specific enthalpy (enthalpy per particle of the accreting fluid) at spatial infinity: $\dot{\mathcal{M}}=-\alpha\,h_\infty$ where $\alpha$ denotes a constant (positive for ordinary matter and negative for phantom matter). If the time scale of accretion $t$ is such that $t\ll \tau\equiv \mathcal{M}/(|\alpha|h_\infty)$, then
	\begin{equation}\label{a11}
		\mathcal{M}(t)=\mathcal{M}\Big[1-\text{sgn}(\alpha)\,\frac{t}{\tau}+\mathcal{O}\Big(\frac{t^2}{\tau^2}\Big)\Big],
	\end{equation}
	where $\mathcal{M}$ denotes the initial mass prior to accretion and $\text{sgn}(\alpha)$ denotes the sign of $\alpha$. By~\eqref{a10}, the rate of change of the throat radius follows the same law
	\begin{equation}\label{a12}
		r_0(t)=r_0\Big[1-\text{sgn}(\alpha)\,\frac{t}{\tau}+\mathcal{O}\Big(\frac{t^2}{\tau^2}\Big)\Big],
	\end{equation}
	where $r_0$ denotes the initial throat radius prior to accretion.
	
	\subsection{Thin accretion disk}
	The simplifying assumptions, governing the theory of thin accretion disk and used for solving the equations describing the conservation and radiation laws, are well described in~\cite{Thorne:1974a,Thorne:1974b}. Within these assumptions it is shown that the mass accretion rate $\dot{\mathcal{M}}$ (usually taken of the order of $10^{-12}M_\odot/\text{yr}$~\cite{Chen:2011,Chen:2012,Zhu:2021} or $2.5\times 10^{-5}M_\odot/\text{yr}$~\cite{Lobo:2009}) is constant and proportional to the surface mass density $\Sigma$: $\dot{\mathcal{M}}=-\alpha\,\Sigma$ where $\alpha$ is another constant. If the time scale of accretion $t$ is such that $t\ll \tau\equiv \mathcal{M}/(|\alpha|\Sigma)$, then Eqs.~\eqref{a11} and~\eqref{a12} still apply to thin accretion disks.
	
	In both cases treated here, spherical accretion and thin accretion disk, the constant $\alpha$ depends on the metric of the central object and on the properties of the accreting fluid~\cite{Azreg:2018,Thorne:1974a} as does the time parameters $\tau$.
	\begin{figure*}
		\centering
		\includegraphics[width=0.6\textwidth]{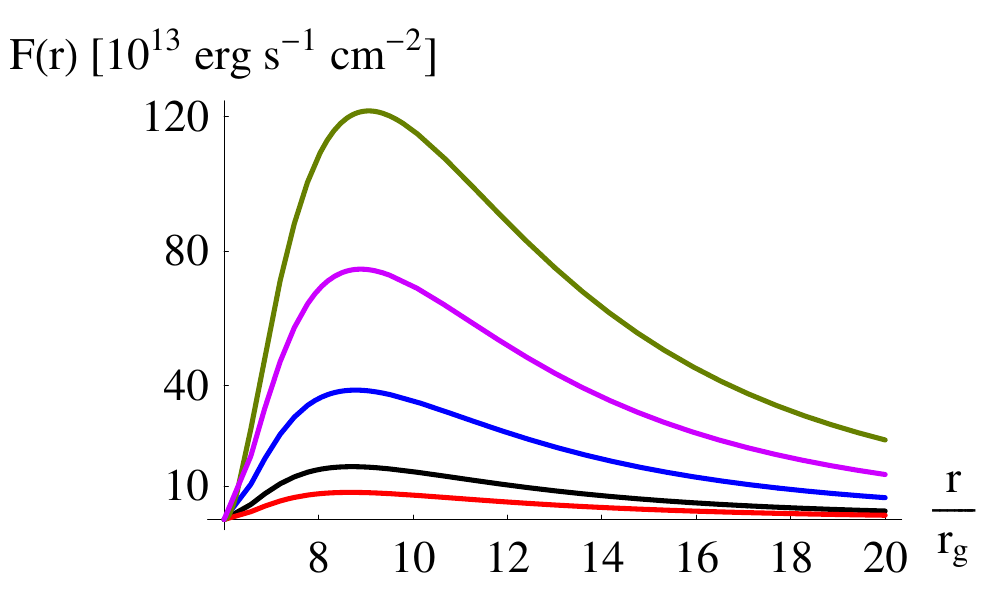}
		\caption{Energy flux $F(r)$ versus $\frac{r}{r_g}$. In these plots we took the mass of the WH model III~\eqref{WHIIIa} $\mathcal{M}=10^6\times M_\odot$, $r_0=kr_g$ [$2<k<6$~\eqref{WHIIIc} and~\eqref{en6}], $M=\frac{4\mathcal{M}}{2k+1}$, $c=299792458$ m/s, $M_\odot=1.9888\times 10^{30}$ kg, and $\text{accretion mass rate }=2.5\times 10^{-5}M_\odot/\text{yr}$. Red Plot: Schwarzschild BH with mass $M_S=10^6\times M_\odot$. Black Plot: WH model III with $k=2.03$. Blue Plot: WH model III with $k=3$. Purple Plot: WH model III with $k=4$. Green Plot: WH model III with $k=5$.\label{Fig-energy}} 
	\end{figure*}
	
	The investigation of circular motion in the plane $\theta =\pi/2$ for the case of a general metric is easily performed using Lagrangian formulation~\cite{Chandra:1983}. Our results are as follows. The specific energy, specific angular momentum and angular velocity of the particle are given by
	\begin{equation}
		\label{en1}E=cA\sqrt{\frac{C'}{AC'-CA'}},\, L=cC\sqrt{\frac{-A'}{AC'-CA'}},\,\Omega=\sqrt{\frac{A'}{C'}},
	\end{equation}
	where the prime notation denotes derivative with respect to $r$ and $c$ is the speed of light. To determine the radius of the marginally stable circular orbit, $r_{\text{ms}}$, one has to solve the equation
	\begin{equation}\label{en2}
		E^2 \Big(\frac{A''}{A^2}-\frac{2 A'^{\,2}}{A^3}\Big)=L^2 \Big(\frac{C''}{C^2}-\frac{2 C'^{\,2}}{C^3}\Big),
	\end{equation}
	which upon using~\eqref{en1} reduces to 
	\begin{equation}\label{en3}
		\frac{A''}{A'}-\frac{2A'}{A}=\frac{C''}{C'}-\frac{2C'}{C}.
	\end{equation}
	This is not a differential equation, rather an algebraic equation once $A(r)$ and $C(r)$ are given. The solution of which yields the radius of the marginally stable circular orbit [for the Schwarzschild BH, the algebraic equation~\eqref{en3} yields the well known value $r_{\text{ms}}=6r_g$]. In the general case Eq.~\eqref{en3} can be solved only numerically. In the case $C=r^2$ we solve it by
	\begin{equation}\label{en4}
		r_{\text{ms}}=\frac{3AA'}{2A'^{\,2}-AA''}\bigg|_{r=r_{\text{ms}}}.	
	\end{equation}
	Now, we use the well known expression $F(r)$ of the radiant energy over the disk~\cite{Thorne:1974a,Lobo:2009}
	\begin{equation}\label{en5}
		F(r)=-\frac{\text{mass rate}}{4\pi\sqrt{|g|}}~\frac{\partial_r \Omega}{(E-\Omega L)^2}~\int_{r_{\text{ms}}}^{r}(E-\Omega L)\partial_r L~dr,
	\end{equation}
	where $g$ is the determinant of the metric. Taking $\text{accretion mass rate }=2.5\times 10^{-5}M_\odot/\text{yr}$, we obtained the plots shown in Fig.~\ref{Fig-energy} for a Schwarzschild BH (red plot) and the WH model III~\eqref{WHIIIa} having the same mass $M_S=\mathcal{M}=10^6\times M_\odot$. For the WH III we have
	\begin{equation}\label{en6}
		r_{\text{ms}}=6r_g.
	\end{equation}  
	This is the same value as for the the Schwarzschild BH because the functions $A(r)$ and $C(r)$ have the same expressions for the WH III and Schwarzschild BH. In the plots we took $M$ [the parameter in~\eqref{mass2}] such that $M=4\mathcal{M}/(2k+1)$ and $r_0=kr_g$ [$2<k<6$~\eqref{WHIIIc}]. It is obvious from Fig.~\ref{Fig-energy} that the energy radiated by the disk accreting onto a WH III may be higher than 10 times the energy radiated by the same disk accreting onto a Schwarzschild BH of same mass as the WH III.
	
	In sketching the radiant energies for our Schwarzschild and wormhole models we assumed that the ADM mass is the physical mass of the central object. For the Schwarzschild model the ADM mass is just the mass parameter in the expression of $g_{tt}$ or $g_{rr}$ while for the wormhole model the corresponding parameter in $g_{tt}$ is nothing but a mere parameter. There are even some wormhole solutions for which $g_{tt}$ is constant with no parameter dependence.
	
	\section{Shadow and infalling gas surrounding a WH}
	\label{secshad}
	Starting from the HJ equation and using the two constants of motion we can easily obtain the geodesic equation of light 
	\cite{Shaikh:2018kfv}
	\begin{eqnarray}
		\frac{dt}{d\lambda} &=& \frac{\mathcal{E}}{e^{2\Phi(r)}}, \\
		\frac{e^{2\Phi(r)}}{\left(1-\frac{b(r)}{r}\right)^{1/2}}\frac{dr}{d\lambda} &=& \pm \sqrt{R(r)}, \\
		r^2 \frac{d\theta}{d\lambda} &=& \pm \sqrt{\Theta(\theta)}, \\
		\frac{d \phi}{d \lambda} &=& \frac{\mathcal{L}}{r^2\sin^2\theta},
	\end{eqnarray}
	where we have introduced
		\begin{equation}
			R(r) = \mathcal{E}^2- \mathcal{K} \frac{e^{2\Phi(r)}}{r^2},\qquad \Theta(\theta) = \mathcal{K} -\mathcal{L}^2\cot^2\theta,
	\end{equation}
	where $\mathcal{K}$ is the Carter constant. We introduce the following new quantities
	\begin{eqnarray}
		\xi = \frac{\mathcal{L}}{\mathcal{E}},\;\;\eta = \frac{\mathcal{K}}{\mathcal{E}^2}.
	\end{eqnarray}
	
	At this stage, one may use the rescaling $\lambda  \to \lambda \mathcal{E}$, such that we obtain the standard expression for the effective potential $V_{\rm eff}(r)$  given by the following relation
	\begin{equation}
		\left(\frac{dr}{d\lambda}\right)^2 + V_{\rm eff} (r)= 0,
	\end{equation}
	in which we have used
	\begin{equation}
		V_{\rm eff}(r) =-\frac{1}{e^{2\Phi(r)}} \left(1-\frac{b(r)}{r}\right)R(r).
	\end{equation}
	
	Next, we can use the radial part for the geodesic equations along with the last expression to investigate  the shadow of the WH geometry by imposing the following conditions
	\begin{equation}
		R(r)=0,\;\; \frac{dR(r)}{dr} =0 ,\;\;\; \frac{d^2 R(r)}{dr^2} >0.
	\end{equation}
	
	Using the radial part of the geodesic motion one can show the following result \cite{Shaikh:2018kfv}
	\begin{equation*}
		\eta=\frac{r^2}{e^{2\Phi(r)}}\Big|_{r_{\text{ph}}}
	\end{equation*}
	with $r=r_\text{ph}$ labels the radial distance of the light ring.  In the case of WHs one can have an additional contribution form the WH throat, namely, as it was argued in Ref. \cite{Shaikh:2018kfv}, the WH throat acts as the position of unstable circular orbits and hence deciding the boundary of a shadow using
	\begin{equation}
		R(r_0)=0,\;\; \frac{d^2 R(r)}{dr^2}\Big|_{r_0} >0,
	\end{equation}
	here $r_0$ represents the WH throat radius. It follows that
	\begin{equation}
		\eta=\frac{r_0^2}{e^{2\Phi(r_0)}}.
	\end{equation}
	From the observational point of view, or so to say in the observer’s sky, we need to use the the celestial coordinates given by \cite{Shaikh:2018kfv}
	\begin{equation}
		X=\lim_{r \to \infty} \left(- r^2 \sin^2\theta_0 \frac{d\phi}{dr}\right)=-\frac{\xi}{\sin\theta_0},
	\end{equation}
	along with
	\begin{equation}
		Y=\lim_{r \to \infty} \left(r^2  \frac{d\theta}{dr}\right)=\left(\eta- \frac{\xi^2}{\sin^2\theta_0}  \right)^{1/2},
	\end{equation}
	with $\theta_0$ being the inclination angle. Combining these results and expressing the shadow radius via celestial coordinates $(X,Y)$ it follows
	\begin{equation}
		R_s= \sqrt{X^2+Y^2}=\frac{r_0}{e^{\Phi(r_0)}}.
	\end{equation}
	
	If we first consider our WH model I, we find that the shadow is determined by the outer photon ring, i.e., $r_\text{ph}=2r_0$. Therefore, the shadow radius results in
	\begin{eqnarray}
		R_s=2 r_0 e^{\frac{3}{4}}.
	\end{eqnarray} 
	
	Now if the accretion takes place, as we have shown the WH throat can change according to \eqref{a12}, which suggests that the shadow radius should change and be time dependent as follows 
	\begin{eqnarray}
		R_s(t)\simeq 2 r_0\Big[1-\text{sgn}(\alpha)\,\frac{t}{\tau}\Big] e^{\frac{3}{4}}.
	\end{eqnarray}
	
	In particular, this shows that if the accreting matter is phantom matter then the $\text{sgn}(\alpha)<0$, as a result the shadow radius will increase in time i.e.,
	\begin{eqnarray}
		R_s(t)\simeq 2 r_0\Big[1+\alpha\,\frac{t}{\tau}\Big] e^{\frac{3}{4}}.
	\end{eqnarray}
	
	But if the accreting matter is ordinary matter, then $\text{sgn}(\alpha)>0$, as a result the shadow radius will decrease according to the formula 
	\begin{eqnarray}
		R_s(t)\simeq 2 r_0\Big[1-\alpha\,\frac{t}{\tau}\Big] e^{\frac{3}{4}}.
	\end{eqnarray}
	
	It is a worth noting here that this result is different for the BH case, where the BH shadow radius should increase with accreting ordinary matter and decrease by adding phantom matter. This result of course has to do with the fact that in order to have an open WH throat we need phantom energy at the WH throat as shown by theorems of general relativity.
	\begin{figure*}
		\centering
		\includegraphics[scale=0.4]{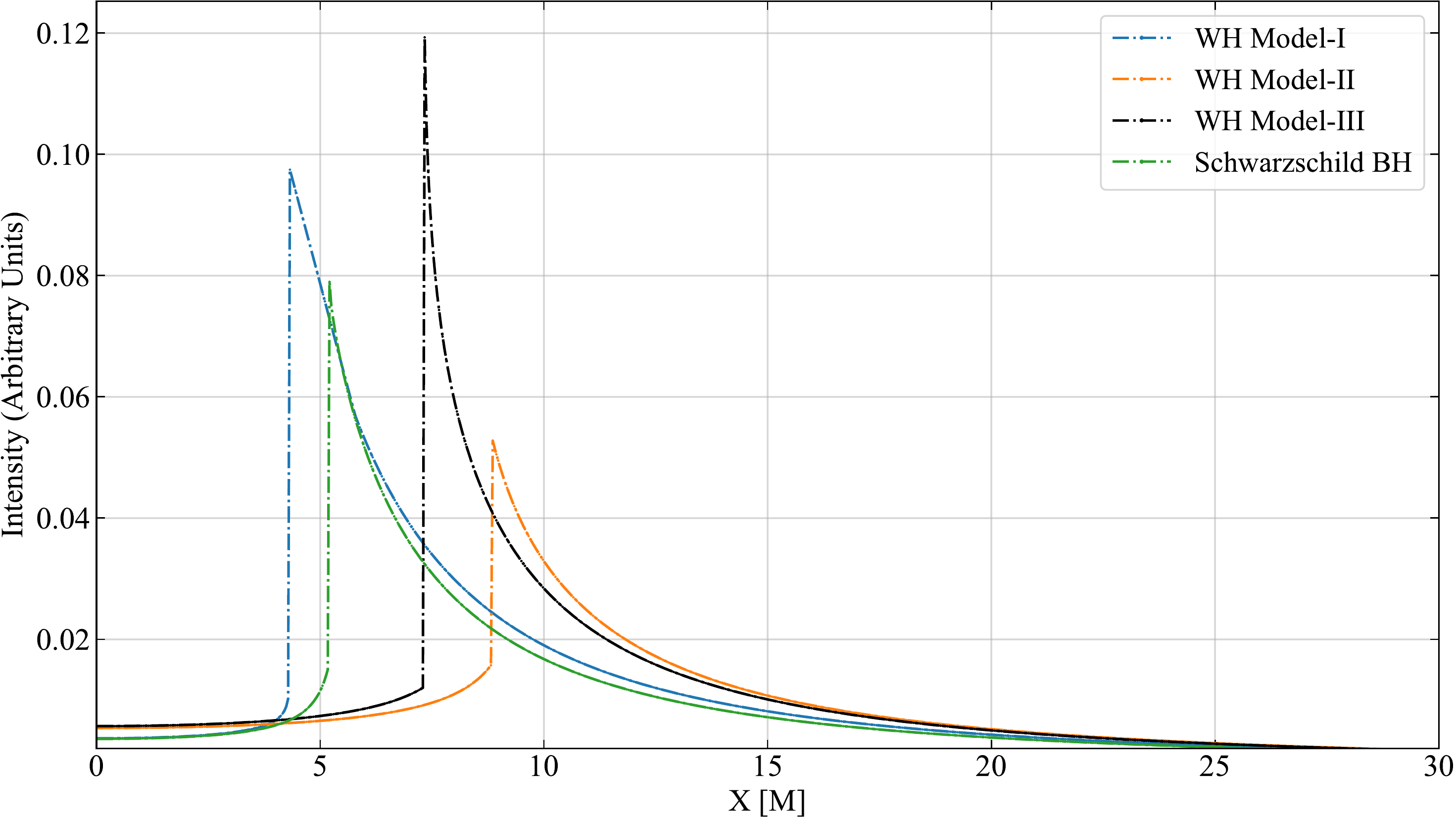}
		\caption{Plots of the corresponding intensities using the infalling gas as seen by a distant observer in different WH geometries and for comparison, the Schwarzschild BH as well. The horizontal axis represents the radial distance of the distant observer from the black hole.}
		\label{figure:intensity}
	\end{figure*}
	
	For WH model II we obtain two solutions for the photon sphere
	\begin{eqnarray}
		r_\text{ph}=\frac{2 Q^2r_0}{Q^2+r_0^2},
	\end{eqnarray}
	and 
	\begin{eqnarray}
		r_\text{ph}=\frac{4 Q^2r_0}{Q^2+r_0^2}.
	\end{eqnarray}
	
	We see that the location of the photon sphere depends on the values of $Q$ and $r_0$. In order to obtain physically acceptable solutions for the photon sphere we should have $r_\text{ph}\geq r_0$. From the motion of S2 star we have obtained the constraint $r_0\sim 7.7$ and $Q\sim 2.06$. This means that the photons spheres are located at $r_\text{ph}=1.028$ and $r_\text{ph}=2.05$, i.e. in the region $r<r_0$. Hence we are left with a contribution only due to the WH throat, since the WH solution is valid only for $r>r_0$.  The WH shadow reads
	
	\begin{equation}
		R_s=\frac{(r_0^2+Q^2)}{r_0^2-Q^2}r_0,
	\end{equation}
	yielding $R_s=8.88$. If the accretion matter is taken into account we can write 
	\begin{equation}
		R_s(t)\simeq \frac{\left(r_0^2\Big[1-\text{sgn}(\alpha)\,\frac{t}{\tau}\Big]^2+Q^2\right)}{r_0^2 \Big[1-\text{sgn}(\alpha)\,\frac{t}{\tau}\Big]^2-Q^2} r_0 \Big[1-\text{sgn}(\alpha)\,\frac{t}{\tau}\Big].
	\end{equation}
	For the WH model III, the photon sphere is located at $r_\text{ph}=3M$, thus, the shadow radius is given by 
	\begin{eqnarray}
		R_s=\frac{r_0}{\sqrt{1-\frac{2M}{r_0}}},
	\end{eqnarray}
	where $r_0 \sim 6.639$ in units of WH mass which is set to unity.  Once the accretion takes place the shadow reads 
	
	\begin{eqnarray}
		R_s(t)\simeq \frac{r_0 \Big[1-\text{sgn}(\alpha)\,\frac{t}{\tau}\Big]}{\sqrt{1-\frac{2M}{r_0 \Big[1-\text{sgn}(\alpha)\,\frac{t}{\tau}\Big]}}}.
	\end{eqnarray}
	In this section, we consider a more realistic model, namely an optically thin, radiating accretion flow surrounding the object and then use a numerical technique (Backward Raytracing) to find the shadow cast by the radiating flow. The calculation of the intensity map of the emitting region requires some assumption about the radiating processes and emission mechanisms. The observed specific intensity $I_{obs}$ at the observed photon frequency $\nu_\text{obs}$ at the point $(X,Y)$ of the observer's image (usually measured in $\text{erg} \text{s}^{-1} \text{cm}^{-2} \text{str}^{-1} \text{Hz}^{-1}$) is given by 
	\begin{eqnarray}
		I_\text{obs}(\nu_\text{obs},X,Y) = \int_{\gamma}\mathrm{g}^3 j(\nu_{e})dl_\text{prop}.
	\end{eqnarray}
	Here we are considering a simplistic case of the accreting gas. We assume that the gas is in radial free fall with a four-velocity $u^{\mu}_e$.
	The four-velocity for the photons $k^{\mu}$ has been found in the previous section. To ease our further calculations, we find a relation between the radial and time component of the four-velocity
		\begin{eqnarray}
			\frac{k^r}{k^t} = \pm A(r) \sqrt{B(r)\bigg(\frac{1}{A(r)}-\frac{b^2}{r^2}\bigg)},
	\end{eqnarray}
	where the sign $+(-)$ is when the photon approaches (towards) or
	away from the massive object.
	The redshift function $\mathrm{g} = \nu_{obs}/\nu_e$ is therefore given by
	\begin{eqnarray}
		\mathrm{g} = \frac{k_{\alpha}u^{\alpha}_o}{k_{\beta}u^{\beta}_e},
	\end{eqnarray}
	where $u^{\mu}_0=(1,0,0,0)$ is the four velocity of the distant observer at infinity.
	For the specific emissivity we assume a simple model in which the emission is monochromatic with emitter's-rest frame frequency $\nu_{\star}$, and we assume that the emission has the simplest radial profile of $r^{-3}$, as is customary done~\cite{em1,em2,em3}:
	\begin{eqnarray}
		j(\nu_{e}) \propto \frac{\delta(\nu_{e}-\nu_{\star})}{r^3},
	\end{eqnarray}
	where $\delta$ is the Dirac delta function. In the simplest case the emissivity at a point on the disk is proportional to the inverse square of the distance from the source times the cosine of the angle at which the ray crosses the disk from the normal, resulting in a form $r^{-3}$ at large radius from the source. The proper length can be written as 
	\begin{eqnarray}
		dl_{\text{prop}} = k_{\alpha}u^{\alpha}_{e}d\lambda = -\frac{k_t}{\mathrm{g}|k^r|}dr.
	\end{eqnarray}
	Integrating the intensity over all the observed frequencies, we obtain the observed flux
	\begin{eqnarray} \label{inten}
		F_\text{obs}(X,Y) \propto -\int_{\gamma} \frac{\mathrm{g}^3 k_t}{r^3k^r}dr.  
	\end{eqnarray}

	The observed flux for different WH geometries and the Schwarzschild BH is plotted in Fig.~\ref{figure:intensity}. We now note the qualitative differences in the shadows and images produced in different models. It can be seen clearly that different geometries have different shadow size and peak intensity. In the case of Model-I Fig.~\ref{figure:shadow} (Top row), although the shadow size is somewhat similar to that of Schwarzschild BH, it can be differentiated. However, this might not be possible in the case of realistic scenario (which we shall discuss in the next section) where we observe the source with radio telescope and has a finite resolution. 
	In the case of Model-II, the size is largest among all the other models we have used and for Model-III the size is in intermediate stage.
	The emissivity and accretion model that we have used in this section to generate the shadow images, provides us with a fine choice of surrounding environment to reconstruct the images with radio observations. We will use these images as model images in the next section.
	
	\begin{table}
		\begin{center}
			\begin{tabular}{|l|l|l|l|}
				\hline
				\textbf{Site} & \textbf{Lat.~($^\circ$)} & \textbf{Long.~($^\circ$)} & \textbf{SEFD. (Jy)} \\
				\hline
				ALMA  & -22.89 & -67.75  & 90    \\
				APEX  & -22.87 & -67.76  & 3500  \\
				GLT   & 72.47  & -38.45  & 10000 \\
				IRAM  & 44.44  & 5.91    & 1500  \\
				JCMT  & 19.7   & -155.48 & 6000  \\
				KP    & 31.78  & -111.61 & 10000 \\
				LMT   & 18.87  & -97.31  & 600   \\
				NOEMA & 44.44  & 5.91    & 700   \\
				SMA   & 19.7   & -155.48 & 4900  \\
				SMT   & 32.53  & -109.89 & 5000  \\
				SPT   & -90.0  & 45.0    & 5000  \\
				\hline
			\end{tabular}
			\caption{Locations of Existing Sites (2021) in the Event Horizon Telescope array configuration.}
			\label{table:1}
		\end{center}
	\end{table}

	\begin{figure}[!htb]
		\centering
		\includegraphics[width=0.48\textwidth]{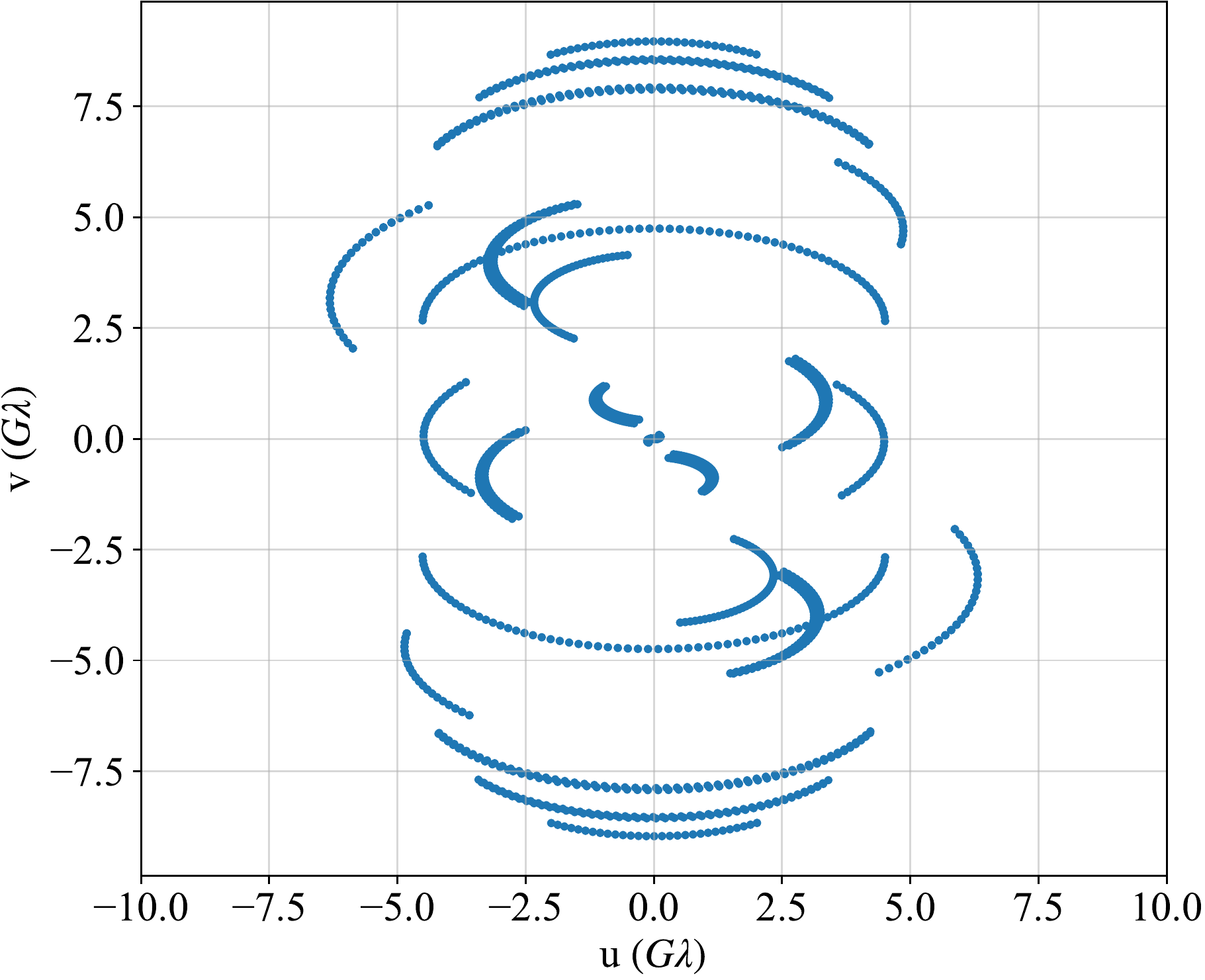}
		\includegraphics[scale=0.48]{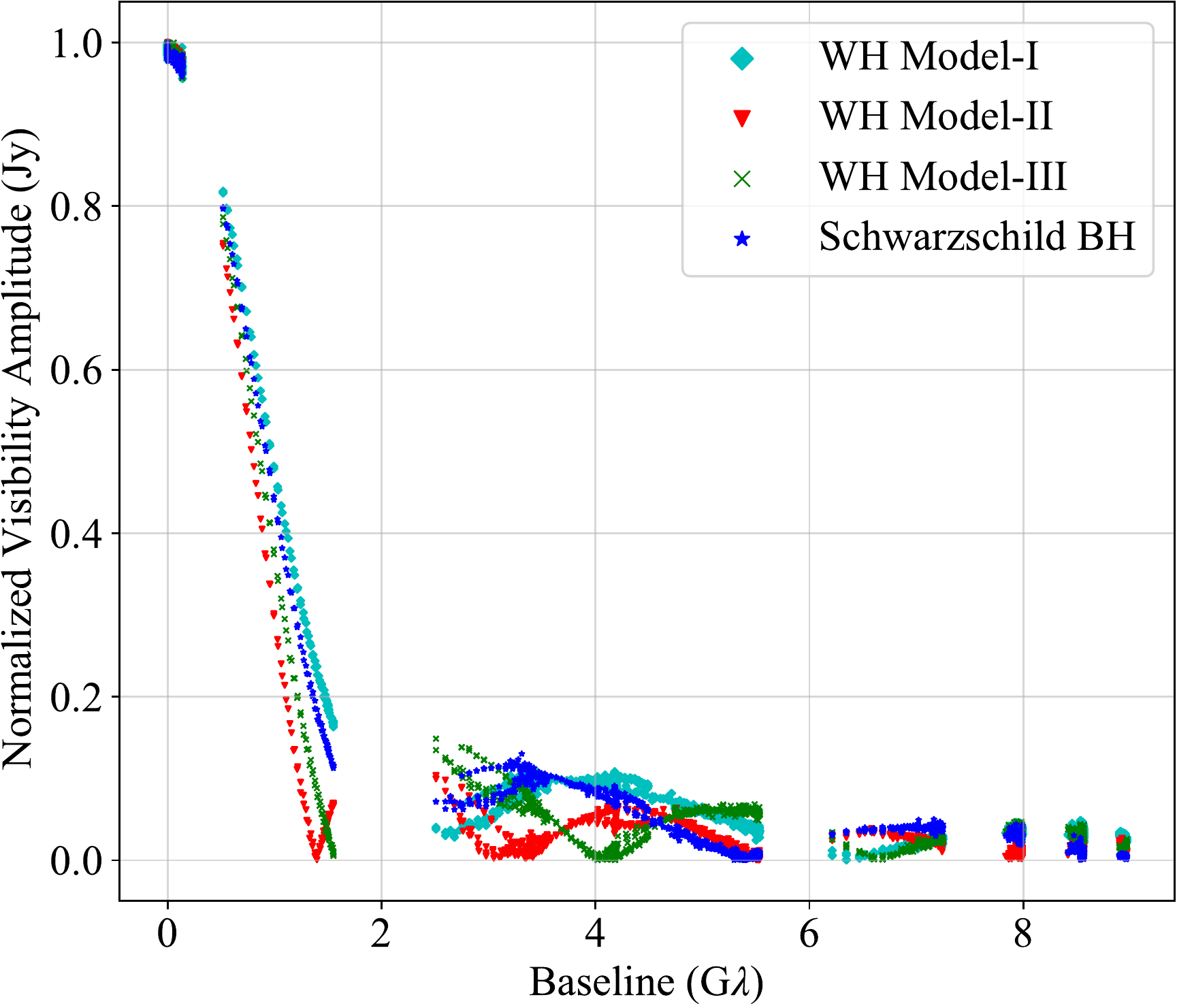}
		\caption{Upper panel: U-V coverage for the EHT-2021 array configuration. Lower panel: Normalized visibility amplitudes for the observations made for different WH geometries: Case I, Case II and Schwarzschild BH respectively. }
		\label{figure:visamp}
	\end{figure}

	\begin{figure*}[!htb]
		\centering
		\includegraphics[width=0.99\textwidth]{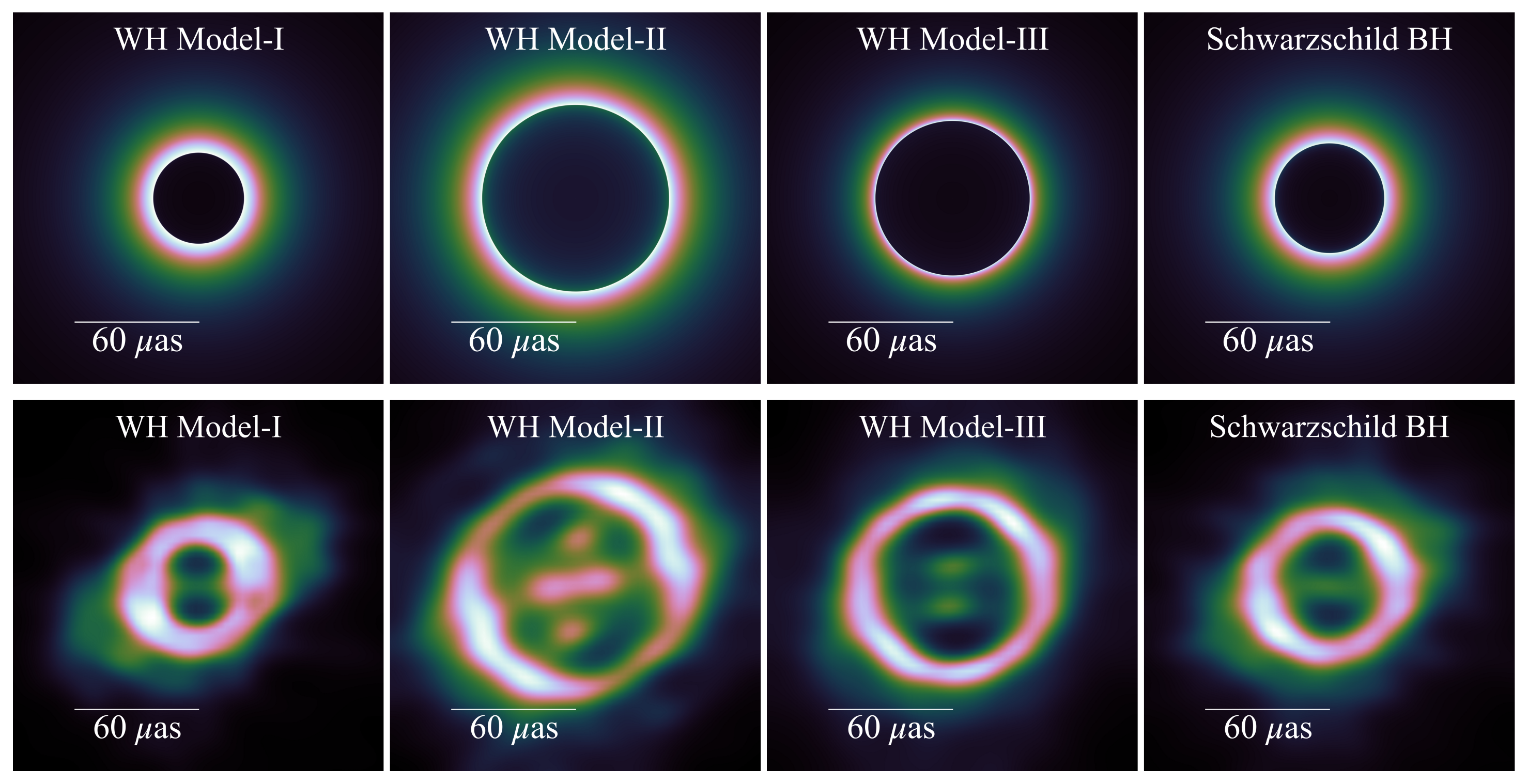}
		\caption{Top row: Ray-traced images of shadows using the radially infalling gas model seen by a distant observer for three WH geometries and Schwarzschild BH respectively. Bottom row: Reconstructed images of the model images in the first row using EHT-2021 array configuration. Read text for more details.}
		\label{figure:shadow}
	\end{figure*}

	\section{Application to the EHT synthetic observations}
	\label{secapp}
	In the previous sections, we generated model shadow images for different WH geometries. While these shadow images have an \textit{ infinite} resolution, in reality these images are mitigated in quality due to interstellar scattering, atmospheric aberration, telescopic limitations etc.  To understand how we might see these sources when the resolution in \textit{finite} and have an Earth-sized radio telescope to observe it, we produce synthetic datasets by keeping the images produced in the previous section as source models and sample them over a particular choice of EHT array configuration. 
	
	Till the April 2018 observing campaign, the EHT had 8 telescopes for observation (\cite{Palumbo1:2019}): the Atacama Large (sub) Millimeter Array (ALMA), in Chile; the Atacama Pathfinder Experiment Telescope (APEX), also in Chile; the James Clark Maxwell Telescope (JCMT), near the summit of Mauna Kea in Hawaii; the Large Millimeter Telescope (LMT), in Mexico; the 30 m telescope on Pico Veleta in Spain (PV); the Submillimeter Array (SMA), located near JCMT; the Submillimeter Telescope (SMT), located on Mount Graham in Arizona; and the South Pole Telescope (SPT), operating at the National Science Foundation’s South Pole research station. In April 2021, 3 more stations were added in the observation campaign \cite{ngeht:2021} (array details can be found in Table \ref{table:1}). 
	
	The synthetic radio images have been generated using the \texttt{ehtim} packaged \cite{Chael:2018}. The following parameters were used in the simulations: $\Delta \nu=4$ GHz bandwidth, $t=24$ hours, corresponding to a full day, at a central frequency of 230 GHz. We closely follow the routine which was used in \cite{Nampalliwar:2021tyz} for Sgr A$^\star$. As the total observing time is an important parameter in the imaging process, this unusually long observation time allows us to present the best-case scenario. This will allow us to properly fill the ($u, v$) plane, which is demonstrated in Fig.~\ref{figure:visamp}. Following this procedure, we perform the synthetic observations of the Galactic center. The amplitudes of visibility are calculated by Fourier transforming the images and sampling them with the array's projected baselines. To simulate realistic observations, we include the effects of thermal noise and phase errors in the simulations. This results in a very detailed visibility amplitude plots, as shown in Fig.~\ref{figure:visamp} (Lower panel).
	
	The major outcome of this whole analysis are the reconstructed images. These images are shown in the second row of Fig.~ \ref{figure:shadow}, meanwhile in the first row we show the corresponding model images for different WH geometries and the Schwarzschild case. The visual inspection of the reconstructed images clearly shows some differences among the three cases based on the shadow size. Case-1 has the smallest size among the chosen models, hence is different in the reconstruction as well, whereas Case-2 has the largest size and has mirrored-crescent like structures. Case-3 is intermediate with a symmetric thin-ring like structure. It is also worth noting that Case-1 to some extent mimics the Schwarzschild case which can be inferred from the fact that in the raytraced images the structure size is almost similar (although the ring is a little thicker in the case of WH Model-1).
	
	It is worth mentioning that in accretion processes the fluid particles' speed approaches the speed of light in the vicinity of the attracting massive center (black hole or wormhole). If the center is a wormhole, the accreting matter does not accumulate in the vicinity of the throat, rather it crosses it to the other sheet of the wormhole. Thus, the accreting matter disappears behind the throat, as does the accreting matter behind the horizon of a black hole, resulting in no changes in the visual appearance.

	\section{Conclusions}
	\label{secconc}
	In this work we have performed a detailed analyses and studied the the possibility of having WHs as a BH mimickers in the galactic center by investigating the dynamics of S2 star and the shadow images.  Firstly, we have constrained their geometry using the motion of S2 star and then reconstructed their shadow images. We have analyzed three WH models, including WHs in Einstein theory, brane-world gravity, and Einstein-Dirac-Maxwell theory. It is argued that, in principle, we can distinguish BHs from WHs since the shadow radius of such WHs depend on the specific model and can be higher or smaller compared to the BH case. We have also considered the accretion of infalling gas model and study the accretion rate and the intensity of the electromagnetic radiation as well as the shadow images. We have shown that the energy radiated by the disk accreting onto a WH may be higher than 10 times the energy radiated by the same disk accreting onto a Schwarzschild BH of same mass as the WH.
	
	For realistic observations, we had to include further effects such as the those pertaining to thermal noise and phase errors in the simulations. We have used the infalling gas model and produced synthetic datasets and obtained interesting reconstructed images. We found that, depending on the WH model, the WH geometry can mimic the BH, however, in principle, one can distinguish these objects by means of the size of the shadow radius. Using each fitting model we have used the reduced chi-squared to test each model and found that WH geometry can mimic very well the BH geometry. On the other hand, we have found that such geometries can be distinguished from the size of the shadow radius.  In our setup, we have used the best fit parameters for the WH geometry obtained from the S2 orbit. For a sufficiently large amount of matter we have shown that the WH throat can change when the accretion of matter takes place, this, on the other hand, suggested that the shadow radius may change with time. Whether the shadow radius  will increase or decrease this depends on the accreting matter. For normal matter the shadow radius of WHs decreases in time, and only increases for phantom matter. This is very different for BH spacetime. 
	We have not considered the effect of matter on the other side of the wormhole. In that situation, one has to study the gravitational effect which propagates through the wormhole and that in principle can effect the motion of objects in our side such as the S2 star orbit (see, for example \cite{d1,d2,d3}). This can imply a different constraints on the wormhole geometries using the S2 star. In our work, we took into account the contribution of the light emitted from infalling particles onto the wormhole throat as seen from our side. But it is an interesting problem to study the effect of matter as well as the gravitational effect from the other side on the optical appearance of the wormholes. This is outside the scope of the present work.
	For the wormhole models II and III, we found from the constraints that $r_0$ is larger than the marginally stable orbit which has to be excluded in the integration domain. This is one of the reasons why we used infalling gas model in our galactic center to reconstruct the images and not the accretion disk model. However, in general, the constraints on the wormhole throat can depend on the type of the galactic center and different galaxies may yield different values for $r_0$. For active galactic nuclei we do expect an accretion disk to be present with the throat radius smaller than the marginally stable orbit, while a possible constraint using combinations of observations, showing the opposite, may rule out some of the wormhole models. Our galactic center is inactive galactic nucleus and we can not easily rule out models II and III.
	Another issue are the astrophysical uncertainties related to the visual size of the shadow using the accretion disk. In general there are many accretion models, but it is interesting that the photon ring is a universal quantity. We also expect, these uncertainties to depend on the peculiar  accretion model.
	In fact, we plan to explore in the near future in more details the variation of the shadow radius due to the accretion of matter in the sense that it might be a useful tool in the future to distinguish BHs from WHs using astrophysical observations.\\

	\subsection*{Acknowledgements}
	The work of QW and MJ is supported in part by the National Key Research and Development Program of China Grant No.2020YFC2201503, the Zhejiang Provincial Natural Science Foundation of China under Grant No. LR21A050001, the Zhejiang Provincial Natural Science Foundation of China under Grant No.LY20A050002, the Fundamental Research Funds for the Provincial Universities of Zhejiang in China under Grant No. RF-A2019015, and National Natural Science Foundation of China under Grant No. 11675143.

\end{document}